\documentclass[a4paper,fleqn]{cas-dc}
\DeclareUnicodeCharacter{2212}{-}
\usepackage{enumitem}
\usepackage{subcaption}
\usepackage[authoryear]{natbib}

\def\tsc#1{\csdef{#1}{\textsc{\lowercase{#1}}\xspace}}
\tsc{WGM}
\tsc{QE}
\tsc{EP}
\tsc{PMS}
\tsc{BEC}
\tsc{DE}

\begin{document}
\let\WriteBookmarks\relax
\def\floatpagepagefraction{1}
\def\textpagefraction{.001}
\renewcommand{\printorcid}{} 
\shorttitle{Extreme Resolution Brain Registration}

\shortauthors{Nazib et~al.}

\title [mode = title]{An Adversarial Approach to Register Extreme Resolution Tissue Cleared 3D Brain Images}                      



%
\author[1,2]{Abdullah Nazib}




\author[2]{Clinton Fookes}

\author[1,3]{Dimitri Perrin}
\cormark[1]
\ead{dimitri.perrin@qut.edu.au}

\affiliation[1]{organization={School of Computer Science, Queensland University of Technology},
    city={Brisbane},
    state={QLD},
    country={Australia}}

\affiliation[2]{organization={School of Electrical Engineering and Robotics, Queensland University of Technology},
    city={Brisbane},
    state={QLD},
    country={Australia}}

\affiliation[3]{organization={Centre for Data Science, Queensland University of Technology},
    city={Brisbane},
    state={QLD},
    country={Australia}}

\cortext[cor1]{Corresponding author}


\begin{abstract}
We developed a generative patch based 3D image registration model that can register very high resolution images obtained from a biochemical process name tissue clearing. Tissue clearing process removes lipids and fats from the tissue and make the tissue transparent. When cleared tissues are imaged with Light-sheet fluorescent microscopy, the resulting images give a clear window to the cellular activities and dynamics inside the tissue.Thus the images obtained are very rich with cellular information and hence their resolution is extremely high (eg .$2560\times2160\times676$). Analyzing images with such high resolution is a difficult task for any image analysis pipeline.Image registration is a common step in image analysis pipeline when comparison between images are required. Traditional image registration methods fail to register images with such extant. In this paper we addressed this very high resolution image 
registration issue by proposing a patch-based generative network named InvGAN. Our proposed network can register very high resolution tissue cleared images. The tissue cleared dataset used in this paper are obtained from a tissue clearing protocol named CUBIC. We compared our method both with traditional and deep-learning based registration methods.Two different versions of CUBIC dataset are used, representing two different resolutions 25\% and 100\% respectively.
Experiments on two different resolutions clearly show the impact of  resolution on the registration quality. At 25\% resolution, our method achieves comparable registration accuracy with very short time (7 minutes approximately). At 100\% resolution,
most of the traditional registration methods fail except Elastix registration tool.Elastix takes 28 hours to register where proposed InvGAN takes only 10 minutes.
\end{abstract}




\maketitle

\section{Introduction}
In medical applications, image registration is a fundamental step to check correspondence between images. 
Voxel-to-voxel or pixel-to-pixel correspondence is required to analyze tissue differences, changes of tissues, organs or tumors over time, etc.
Image registration is traditionally addressed as an optimization problem. 
However, iteratively optimizing alignment parameters by an objective function requires lot of computation specially when deformable registration is required.
The number of parameters and the associated computational cost increase with the resolution and dimensions of the images. 
This makes such approaches impractical for high-resolution images.

Recently, a biochemical process called tissue clearing has emerged. It is used to remove light-obstructing elements from soft tissues and enable biologists to take 3D single-cell resolution images without sectioning.
A number of tissue clearing methods have been developed, including BABB \citep{Dodt2007}, Scale 
\citep{Hama2011}, SeeDB \citep{Ke2013}, CLARITY \citep{Chung2013}, iDISCO \citep{Renier2014} and CUBIC \citep{Susaki2014}.
Irrespective of the clearing protocol, light-sheet fluorescence microscopy (LSFM) is then used to produce images of entire organs with a resolution of a few micrometres.
For instance, the images from the CUBIC dataset used in this paper are about three orders of magnitude larger than typical MRI images ($6.45\times6.45\times10\mu m^3$ as opposed to $(0.86\times0.86\times1.5mm^3)$.

While this level of detail is crucial from a biological point of view, it makes it difficult to use conventional registration methods for these images, due to their very long computation time and consumption of large compute resources.   
Recently, deep-learning (DL) based methods has achieved remarkable success in image registration \citep{Balakrishnan2018,Eppenhof2019,Cao2018}.
Very high computational efficiency and accuracy makes these methods a favorable choice over the iterative methods.
However, training a DL-based registration method requires large amount of data. 
In our case, having large amount of tissue cleared data (similar to \citet{Balakrishnan2018}) is practically impossible.
Another limitation is that deep-learning based methods have to estimate a large number of deformable parameters, and this number of registration parameters increases with the resolution.
This makes such approaches very demanding, for both memory and computating power.

Based on our observation, we set a list of criteria for a DL-based registration method suited to the tissue-clearing context:

\begin{itemize}
    \item The learning framework has to be trainable with small amount of data.
    \item It has to be able to take high-resolution images with limited computational resources.
    \item It has to be scalable.
    \item It should not be dependent on specific reference image; i.e., any arbitrary image pair can be registered without further training.   
\end{itemize}

In this paper, we propose a generative inverse-consistent method (InvGAN) for the registration of images obtained with tissue clearing and LSFM, aimed at addressing these criteria.
The proposed method simultaneously generates forward and backward deformation fields. 
Two discriminator networks compare the registration quality of both the forward- and backward-transformed images with the corresponding target images. 

The LSFM images contain cellular information and the intensities are not continuous but discrete. 
The difference of cellular structure between two brains is also small. 
Hence, the deformation in these images is small and smooth.
Comparing the target image with the transformed source image by using only intensity based image metric would not provide information on small local changes in deformations.
A key difference of our approach compared to prior work is the use of distribution matching using adversarial loss. 
Using two discriminators for both forward and backward deformation field matches the distribution of images and gives a more realistic and smooth flow.
Our key contributions are:
\begin{enumerate}[label=(\alph*)]
    \item A patch-based registration model that can handle whole-organ images at a single-cell resolution. We tested our model with images at 100\% resolution following CUBIC-based clearing and LSFM imaging. 
    To the best of our knowledge, no deep-learning based registration method tried to register images at that scale.  
    \item The use of multiple decoders and their corresponding discriminators ensures that they generate smooth and realistic flow vectors suitable for LSFM data.
    We conducted landmark registration to measure how realistic the flow vectors from our method are and compare them with baseline methods.
    \item For a patch-based registration model, patch effects can significantly reduce the quality of the registered images.
    The use of adversarial loss reduces the patch effects in the images.
    We also conducted experiments without adversarial loss and result shows the impact of adversarial loss on the registered images. 
\end{enumerate}

\subsection{Related Work}
Deep neural networks have recently been successfully used in medical image registration, based on supervised \citep{Yang2016, Rohe2017, Sokooti2017} or unsupervised methods \citep{Li2018, Balakrishnan2018}.
The development of these deep learning based registration methods open the door to apply adversarial training for medical image registration. 
\citet{Yan2018} appear to be the first to propose an adversarial method for medical image registration.
In that paper, two similar convolution networks are used as generator and discriminator.
The last layer of both networks is a fully connected layer.
In the generator network, the last fully-connected layer regress the transformation parameters while the last fully connected layer in discriminator provides the similarity score.
The generator network takes magnetic resonance (MR) and transrectal ultrasound (TRUS) images as input and estimates transformation parameters which are used by an image resampler to resample the moving image.
The discriminator takes two pairs of images to discriminate the quality of the registration.
For the discriminator training, a well-aligned image pair is used as ground-truth alignment and the discriminator determines the quality of the alignment/transformation parameters estimated by the generator.
In this work, Wasserstein loss \citep{Arjovsky2017} is used to train the discriminator network while the generator network is trained with adversarial loss along with $l1$ loss between estimated flow and a randomly generated flow.
The GAN framework is trained and tested on a MR/TRUS dataset with 763 pairs of images.
The performance is measured using Dice score and no baseline method is used to compare the performances. Though, the performance of the GAN framework proposed in this paper is not compared to other tools, and it is therefore difficult to asses it, the method demonstrates the applicability of adversarial training for image registration.
Again, the dependence on an already registered image pair for the discriminator training is a major limitation of this method.

\citet{Mahapatra2018a} proposes another GAN-based image registration framework. 
In this GAN framework, the training strategy of cycleGAN \citep{Zhu2018} is used, with cyclic loss.
The generator network takes the reference and moving images as input and generates a transformed image and transformation parameters.
The discriminator differentiates between the transformed image and reference image along with the difference between reference flow and estimated flow by generator network.
There is no use of Spatial Transformer Network (STN) like the previous method, since the generator network directly generates the moving image.
During the training, the method uses normalized mutual information loss (NMI), structured similarity index (SSIM) loss, and VGG loss to train the generator network.
The NMI loss maintains the intensity distribution of the transformed image similar to  moving image while SSIM loss maintains the structural similarity between the transformed image and reference image.
To make the deformation field consistent and to maintain its reversibility the cyclic loss similar to cycleGAN \citep{Zhu2018} is used.
Training this network begins with the training of RESNET.
The weights from the trained RESNET are used to initialize the generator network.
A retinal fundus dataset and a cardiac dataset are used to evaluate performance and compared with two base line methods, DIRNET \citep{DeVos2017} and Elastix \citep{Klein2010a}. 
For both datasets, it is found that the GAN based model without cyclic loss achieved the best performance.
However, the method still has limitations.
Firstly, the dependence on VGG layers is a weakness and authors did not mention the performance of the network without VGG layers in the loss function.
Secondly, the application of cycle consistency requires the generator networks to be used twice during training, which makes the training difficult in an ordinary computing environment.
Moreover, the performance tables show that the model actually performs better without cycle consistency.

\citet{Fan2018} proposed a patch-based image registration method consisting of a generator and a discriminator network.
The generator is a U-net architecture \citep{Ronneberger2015} and the discriminator network has a fully connected regression layer with one unit in the output which provides the alignment score from 0 to 1.
During training, the generator takes a pair of corresponding image patches with size $64\times64\times64$ from both the reference image and moving image.
It generates a deformation field of size $24\times24\times24$ to warp the moving image into the space of target image.
Like \citet{Yan2018} the discriminator takes two pairs as input.
From a pair of images previously aligned using another registration software, corresponding patches are extracted and the discriminator treats them as true data/pair.
On the other hand, the pair of target image-patch and transformed moving-image patch is treated as negative or fake image pair/data.
Only adversarial loss is used to train the generator, and to regularize the deformation field a diffusion regularization is used as in \citep{Balakrishnan2018}.
The network is trained and tested on 4 publicly available MR datasets and shows promising results in each of them.
Similar to \citet{Zhu2018}, this patch-based adversarial network is also dependent on the aligned images of other registration tools.
Despite promising performance, the method has some limitations.
Firstly, the network is trained and tested on image patches. From our experience, we found that patch selection for training and testing impacts on the performance of the network. The authors did not mention clearly how they selected patches for training or testing.
Secondly, the authors did not mention how they generated the positive pairs for the discriminator training clearly.
It seems they used other registration software tool but no method name is discussed in the paper.
Thirdly, from our experience on patch based training, the patch by patch flow generation has discontinuity.
In this paper the authors did not mention how they resolved the issue of patch discontinuity. 

\section{Materials and Method}
\subsection{Network Architecture}
\begin{figure}
 	\centering
 	\includegraphics[clip,trim={00mm 10mm 00mm 00mm},width=\linewidth]{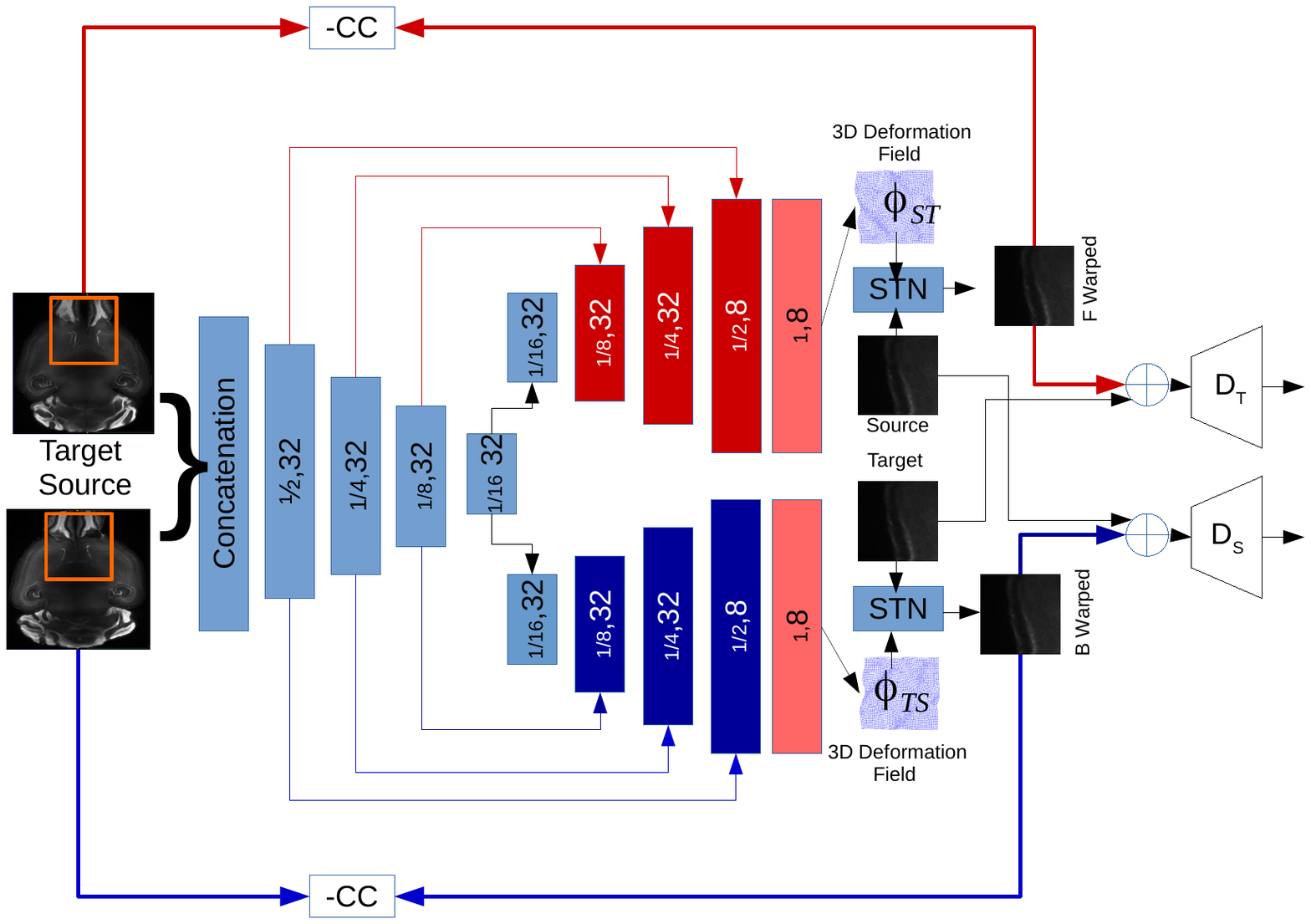}
 	\caption{Inverse Consistent Adversarial Network with two discriminators $D_S$ and  $D_T$, where $D_S$ compares the target image patch with the transformed source patch, and $D_T$ compares the source image patch with the transformed target patch.}
 	\label{fig:Inv_adv}
 \end{figure}

The architecture of our proposed network is shown in Figure~\ref{fig:Inv_adv}.
The network consists of an encoder like VoxelMorph (VM) U-net \citep{Ronneberger2015}, and two decoders.
The input to the network is a concatenation of source and target images.
The encoder branch has four layers, each of which is a convolution layer of $3\times3\times3$ kernel with stride 2 and 32 output channel. 
Among the two decoder branches, one branch is responsible for forward flow computation and the other one is for backward flow computation.
The forward-decoder has one simple convolution layer, three forward computation blocks (red blocks in Figure~\ref{fig:Inv_adv}) with 32, 32 and 8 channels, and two additional convolution layers with 8 and 3 channels.
Each forward computation block contains an upsampling layer, an addition layer and concatenation layer.
The addition layers adds the same sized features from the encoder branch and then concatenates them.
The architecture of the backward-decoder is exactly the same as the forward-decoder with the exception of subtraction layers in the backward computation blocks (blue blocks in Figure~\ref{fig:Inv_adv}).

Each discriminator has six 3D convolution layers with stride 2 and LeakyRelu activation.
The last layer of convolution has $1\times1\times1$ convolution filters.
Each convolution layer reduces the resolution of the input image patch by half.
The last layer of the discriminators produces a single unit output which represents the similarity between input images.

\subsection{Discriminator Loss}
 In our GAN framework, the two discriminators are termed as $D_{S}$ and $D_{T}$ respectively. 
 During training $D_T$ takes the target image as the real image and the transformed source image (with forward transformation) produced by generator network for comparison.
 The forward branch of the generator network produces a forward transformation which is used to warp the source image patch.
 The target discriminator $D_T$ compares the difference between the original target image and the transformed source image using binary cross entropy loss.
 $D_T$ tries to establish whether the image is from generator or it is real.
 The objective function for the target discriminator network is given by:
 \begin{equation}
 \label{ch6:eqn:dis_loss_1}
  \mathcal{L}_{D_T} =\dfrac{1}{N}\sum\mathcal{L_{D_T}}(P_T,1)+\dfrac{1}{N}\sum\mathcal{L_{D_T}}(P_{S\circ\phi},0)
 \end{equation}
 where $\mathcal{L_{D_T}}$ is the binary-cross entropy loss.
 To optimize the generator network to learn a more accurate forward flow, the adversarial loss from the discriminator is used.
 For the generator, the forward adversarial loss is:
 \begin{equation}
\mathcal{L}_{INV_{adv_{D_{T}}}} =\dfrac{1}{N}\sum\mathcal{L_{D_T}}(P_{S\circ\phi},1)
 \end{equation}
 Similar to the target discriminator $D_T$, the source discriminator $D_S$ is also trained with binary cross entropy loss.
 The objective of the source discriminator is to differentiate the difference between the original source image and the transformed target image, which is in the source image space.
 Therefore, the loss for the source discriminator is:
 \begin{equation}
 \label{ch6:eqn:dis_loss_2}
 \mathcal{L}_{D_S} =\dfrac{1}{N}\sum\mathcal{L_{D_S}}(P_S,1)+\dfrac{1}{N}\sum\mathcal{L_{D_S}}(P_{T\circ\phi},0)
 \end{equation}
 Again, the generator needs to be optimized to generate better a backward flow with the adversarial gain learned from the source discriminator.
 The adversarial loss in this case is:
 \begin{equation}
  \mathcal{L}_{INV_{adv_{D_{S}}}} =\dfrac{1}{N}\sum\mathcal{L_{D_S}}(P_{T\circ\phi},1)
 \end{equation}
 
 \subsection{Generator Loss}
 The generator network is optimized with the two adversarial losses.
 The adversarial losses help the generator network learn the distribution of the flow vectors from the input images.
 At the beginning of the adversarial training, the generator sometimes fails to win over the discriminator, and ongoing failing of the generator network makes the training unstable.
 To avoid unstable training, the generator needs an independent loss function that will slowly drive the generator to the Nash equilibrium of the minmax game.
 To drive the generator, we use cross-correlation loss as image similarity metric and cycle loss to ensure reversibility.
 The overall loss function is then given by:
 
 \begin{equation}
 \label{ch6:eqn:generator_loss}
 \mathcal{L}_{Loss} = \mathcal{L}_{similarity}+ \mathcal{L}_{cycle} +\lambda(\mathcal{L}_{INV_{adv_{D_{S}}}}+\mathcal{L}_{INV_{adv_{D_{T}}}})
 \end{equation}
 where
 \begin{equation}
 \mathcal{L}_{similarity} = -CC(S \circ \phi_{ST}, T)-CC(T \circ \phi_{TS},S)
 \end{equation}
 and
 \begin{equation}
 \mathcal{L}_{cycle} =||((T \circ \phi_{TS}) \circ \phi_{ST}) - T||_{1} + ||((S \circ \phi_{ST})\circ \phi_{TS}) - S||_{1}
 \end{equation}

\subsection{Data Preprocessing}
In this paper, we use data from \citet{Susaki2014,Susaki2015}.
Whole brains from Arc-dVenus mice were sampled, cleared using the CUBIC protocol, and imaged using LSFM.
The mouse brains were imaged from two different direction, Dorsal-to-Ventral (D-V) and Ventral-to-Dorsal (V-D).
Here, we use two different resolutions for this data, 25\% and 100\%.  
For the 25\% resolution data, we use the CUBIC informatics protocol \citep{Susaki2015}, in which images are down-sampled to 25\% resolution and then D-V and V-D sides are merged.
20 merged CUBIC brains at 25\% resolution are used to train the network and 3 merged brains are used for testing. 
For the 100\% resolution data, no downsampling operation is used.
Due to memory limitation, the merging operation at 100\% resolution is also discarded.
The network is trained with 20 D-V brains at 100\% resolution and tested with 3 D-V brains at the same resolution.
We chose the D-V direction because the majority of anatomically important regions were more clearly visible in these images than in the V-D direction.

\subsection{Training Patch Selection}
Since training with such high-resolution 3D volumes is difficult due to resource constraints, we employ a patch-based training.
The patch selection itself is a critical step.
The performance of the network depends on the regions from which the patches are selected.
Previously, we used mean intensity of the patches as the selection criteria.
We used 0.2 of the mean patch intensity as the threshold value to decide whether a patch is to be selected or not.
We found that using a specific patch intensity introduced a bias, with most patches selected from only a few specific regions.
This limits the ability of the network to learn deformation parameters for other regions.
To avoid a biased patch selection, we present here a probabilistic patch selection procedure.
We use a step function, shown in Eq.\ref{eq:step}, that gives a weight to a randomly selected patch, based on the mean patch intensity.
The function sets the probability 1 if the mean patch intensity is between 0.1 to 0.35, and if it is more than the given range the probability decreases exponentially.
The mean intensity range given in Eq.\ref{eq:step} is decided empirically.          
\begin{equation}
\label{eq:step}
pdf(x) = \left\{
\begin{array}{ll}
1 & \quad 0.1 <= x <=0.35   \\
\dfrac{10}{e^{K\mu}} & \quad x > 0.2,  K= 6.6\\
0 & \quad otherwise
\end{array}
\right.
\end{equation}

\subsection{Training}
The proposed network is trained and tested at 25\% and 100\% resolution. 
For both cases, there are 20 brains for training and 3 brains for testing.
We use $N\times(N-1)$ combinations of training pairs to train the network.
At 25\% resolution, the image dimension is $640\times540\times169$ and the training patch extraction can therefore be done on the fly.
For each pair of images, 2500 patches are extracted using the \textit{pdf} defined in Eq.\ref{eq:step}. 
Since the image dimension at 100\% resolution is very high ($2560\times2160\times676$), loading two images is difficult.
Patch extraction at this resolution (10,000 patches  from each pair of images) is done before the training starts.
All training images are intensity normalized as well as affine registered to the selected brain (brain-3 of test dataset at both scale) before the training process is applied. 

The network is developed using tensorflow. 
At 25\% resolution, the network is trained and tested in a High-Performance Computing environment with 64 GB RAM, 12 GB Video RAM in Tesla K40m GPU and a single core 2.66GHz 64bit Intel Xeon processor. 
The same configuration is used to train the network at 100\% resolution. 
The data pre-processing and training patch extraction at this resolution is done in a Big-data machine with 4TB memory.


\subsection{Competing Methods}
Five state-of-the art image registration tools have been used in this investigation. Each of these registration tools are tuned and optimized for best performance.
The registration tools selected in our evaluation are based on previous studies like \citep{Klein2010a} and \citep{Xu2016}. We select tools which are automated, easy to use, developed for volume registration and showed consistent performance in previous studies.We exclude tools like FreeSurfer. The FreeSurfer tool is primarily developed for surface registration not for volume. Moreover, the FreeSurfer has a brain labeling algorithm attached with its own labelled atlas which is not suitable for our CUBIC data \citep{Klein2010a}. A brief discussion on each of these tools is as follows:

1) \textbf{IRTK:} One of the early image registration tools by \citep{Rueckert1999} for breast MRI images, using voxelized mutual information similarity and a free-form deformation model. Before starting registration, IRTK applies contrast enhancement to make the similarity measure insensitive to intensity change. A hierarchical transformation model is applied to capture global and local motion of the volume data where global motion captured by affine model and local motion is by a non-linear, free-form deformation model. Voxel-based Normalized mutual information is used as the similarity measure. In our comparison, we followed the same settings as \citep{Xu2016} except for the B-spline control points. Since the pixel spacing of CUBIC dataset is very small, the control point spacing is set to 5mm, which is the highest possible value for this method. The IRTK codes are available in https://github.com/BioMedIA/IRTK.

2) \textbf{Elastix:} 
One of the popular registration tool developed by \citep{Klein2010a}, for CT and MRI images with large set of common registration algorithms. This tool consists of many algorithms for similarity, optimization, regularization, interpolation, transformation etc. For the similarity measure, Elastix includes mutual information (MI), Normalized mutual information (NMI), Cross-correlation (CC), mean squared difference (MSD) etc. The transformation models included in the Elastix library are rigid; affine with different degree of freedom; B-spline with physics based control points in uniform and non-uniform grids; a set of optimization methods namely gradient descent, quasi-Newton, nonlinear conjugate gradient (with several variants); and a number of stochastic gradient descent methods. All these options add flexibility to choose required components whenever necessary. We consider the elastix parameter settings used in \citep{Hammelrath2016} for rigid, affine and nonlinear registration. The Elastix codes are available in https://github.com/SuperElastix/elastix.

3) \textbf{ANTS:} 
Advanced Normalization is developed by \citep{Avants2008},\citep{Avants2011}. ANTS use symmetric diffeomorphic normalization method for non-linear transformation. In ANTS, cross-correlation is maximized in a symmetric diffeomorphic map and uses Eular-Lagrange equations for optimization. The diffeomorphic map preserves the topology map along with invertible transformation parameters and gives sub-pixel accuracy. The parameter settings for the ANTS tool is derived from ANTS example script. In evaluation by \citep{Xu2016}, two different setups of ANTS tool were used with two different similarity metric (Cross-Correlation and Mutual Information), which they considered as two separate methods. In our settings, only cross-correlation is used as a similarity measure; the number of resolution levels is three, with 100 iterations in each sampling level. The ANTS codes are available in https://github.com/ANTsX.

4) \textbf{NiftyReg:} 
A promising registration tool developed by \citep{Modat2010}. It’s an extended version of IRTK, based on free form deformation. In this method, the gradient of normalized mutual information of each B-spline control point is calculated and used in gradient descent-based optimization method. The algorithm was implemented with parallel processing, but in our evaluation only a CPU version is used.  The parameter setting of this tool for CUBIC evaluation
is exactly the same as settings mentioned by \citep{Xu2016}. The number of iterations is 1000 for free-form deformation and 500 intensity threshold for both source and target image. The NiftyReg binary files are downloaded from https://github.com/KCL-BMEIS/niftyreg/wiki.

5) \textbf{VoxelMorph:} This is the first deep-learning based registration method \citep{Balakrishnan2018}. The VoxelMorph integrates a fully convolutional U-net architecture with a Spatial transformer and train them simultaneously. Unlike deep-learning based approaches developed before, this architecture directly takes fixed and moving images in 3D form instead of taking image patches. Using cross-correlation as objective function to estimate dissimilarity between fixed and instantaneously moved moving image, the network is trained in unsupervised manner. A diffusion regularizer is used to prevent training over-fitting of the network.

For our experiment, ANTS, Elastix and IRTK are built from the source code given in their respective code repository while binary files of NiftyReg is downloaded.

\section{Results}
The performance of the proposed method is evaluated on two different resolution scales, 25\% and 100\% respectively. We compared the quantitative performance of the registration methods by normalized cross-correlation and mutual information for test brains. We have only three CUBIC brains to test and for the convenience we note them as brain-1, brain-2 and brain-3. The performance scores showed in tables are measured by making brain-3 as target and other two brains as moving image.   
For qualitative evaluation, we show the same brain slice extracted from all test brains and overlaid on the target brain with different color map. In all of our experiments, the target brain is mapped with red color and registered brains are mapped with green color. The dissimilarity in the cerebellum (posterior part) region is common due to high variability of that part from brain to brain. The visual registration quality is assessed by considering regions like hippocampal formation, dentate gyrus to be similar and aligned. Again, to facilitate the qualitative evaluation process at 25\%, cropped and zoomed patches from hippocampal formation and dentate gyrus regions are also presented.    

\subsection{Results at 25\% Resolution}
In Table~\ref{Tab:25_quantitative}, proposed InvGAN achieves the highest scores in both test brains in terms of cross-correlation score with scores 0.8877 and 0.9627. In mutual-information, our InvGAN still remains one second best with tiny difference with ANTS tool. Elaxtix tool achieves third best results in both metrics. VoxelMorph, the only deep learning based competitor
performs better than conventional NiftyReg and IRTK. The IRTK tool become the least performing 
tool in this experiment. 

\begin{figure}[!htb]
	\begin{center}
	    \begin{minipage}[t]{4cm}
			\includegraphics[width=3.9cm,height=3.9cm]{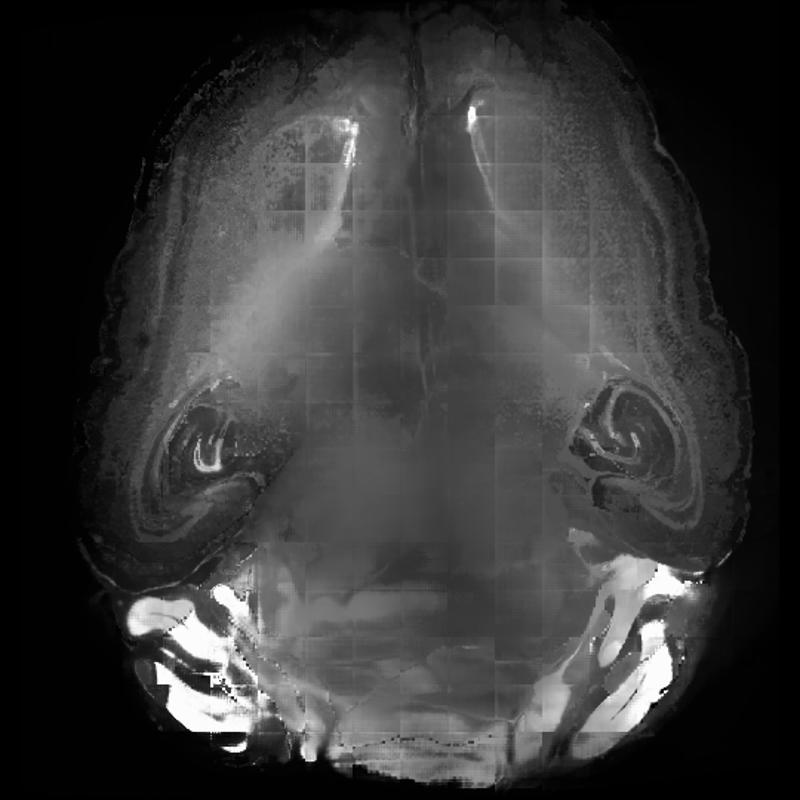}
				\subcaption{no adversarial loss}
		\end{minipage}%
		\begin{minipage}[t]{4cm}
			\includegraphics[width=3.9cm,height=3.9cm]{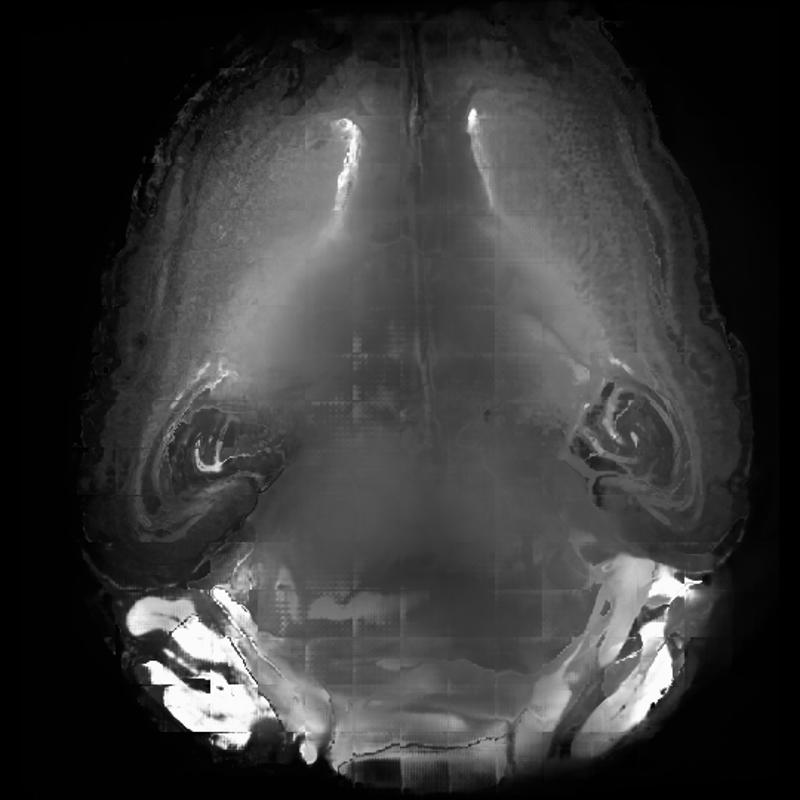}
				\subcaption{with adversarial loss}
		\end{minipage}%
		\caption{Box artifacts in non-Adversarial training of InvGAN}
		\label{Fig:comparison_nonGAN}
	\end{center}
\end{figure}

To further explore the capability of of InvGAN architecture, a non-generative version is trained and tested without discriminator training. The results of the model without adversarial training is also presented in the last row of Table~\ref{Tab:25_quantitative}. The quantitative results of non-adversarial training of is extremely high in our context. To verify the registration quality of non-adversarial training, registered images are checked. In Figure~\ref{Fig:comparison_nonGAN} a comparison between training with and without adversarial loss is presented. It is found that without adversarial loss the image quality drastically affected by the box artifacts from the patch-based training strategy.             

\begin{table}[ht]
	\begin{center}
		\caption{\small  Performance comparison at 25\% Resolution}
		\label{Tab:25_quantitative}
		\small
		\begin{tabular}{l {c}p{0.1\linewidth} {c}p{0.1\linewidth} {c}p{0.1\linewidth} {c}p{0.1\linewidth}}
			\hline
			Methods & brain-1 &brain-1 & brain-2 & brain-2\\
		        &  CC        &MI  &CC &MI\\
		    \hline
			BEFORE                  &0.6763  	&0.9147         &0.8016  	&1.2006\\
			ANTS			    	&0.8712 	&1.3583         &0.9391		&1.5918\\
			Elastix				    &0.8686		&1.1542			&0.9102		&1.2903\\
			NiftyReg			    &0.7166		&1.1319			&0.7570 	&1.1677\\
			IRTK				    &0.6148		&0.8205			&0.6713		&0.9154\\
			VoxelMorph          	&0.7877		&1.0620			&0.9126		&1.4009\\
			InvGAN                  &0.8877     &1.2307          &0.9627   	&1.5452\\
			no-Adversarial loss		&\textbf{0.9239}		&\textbf{1.4328}	&\textbf{0.9712}		&\textbf{1.6708}\\
			\hline
		\end{tabular}
	\end{center}
\end{table}

\begin{figure*}[!htb]
    \centering
    \includegraphics[clip,trim={00mm 20mm 00mm 10mm},width=\textwidth]{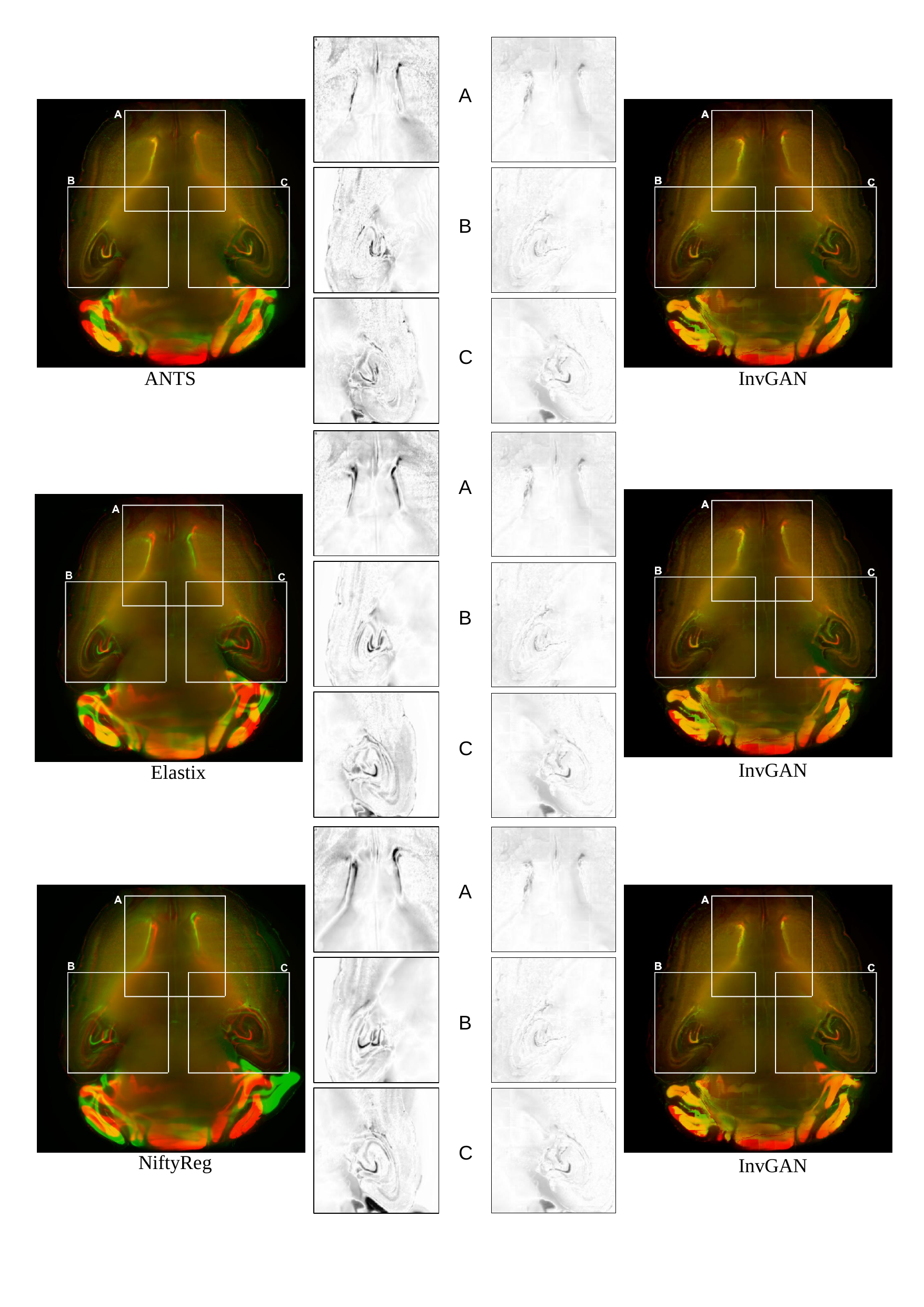}
    \caption{Visual Comparison at 25\% Resolution-1}
    \label{fig:25_compareDiff_1}
\end{figure*}

\begin{figure*}[!htb]
    \centering
    \includegraphics[clip,trim={00mm 150mm 00mm 10mm},width=\textwidth]{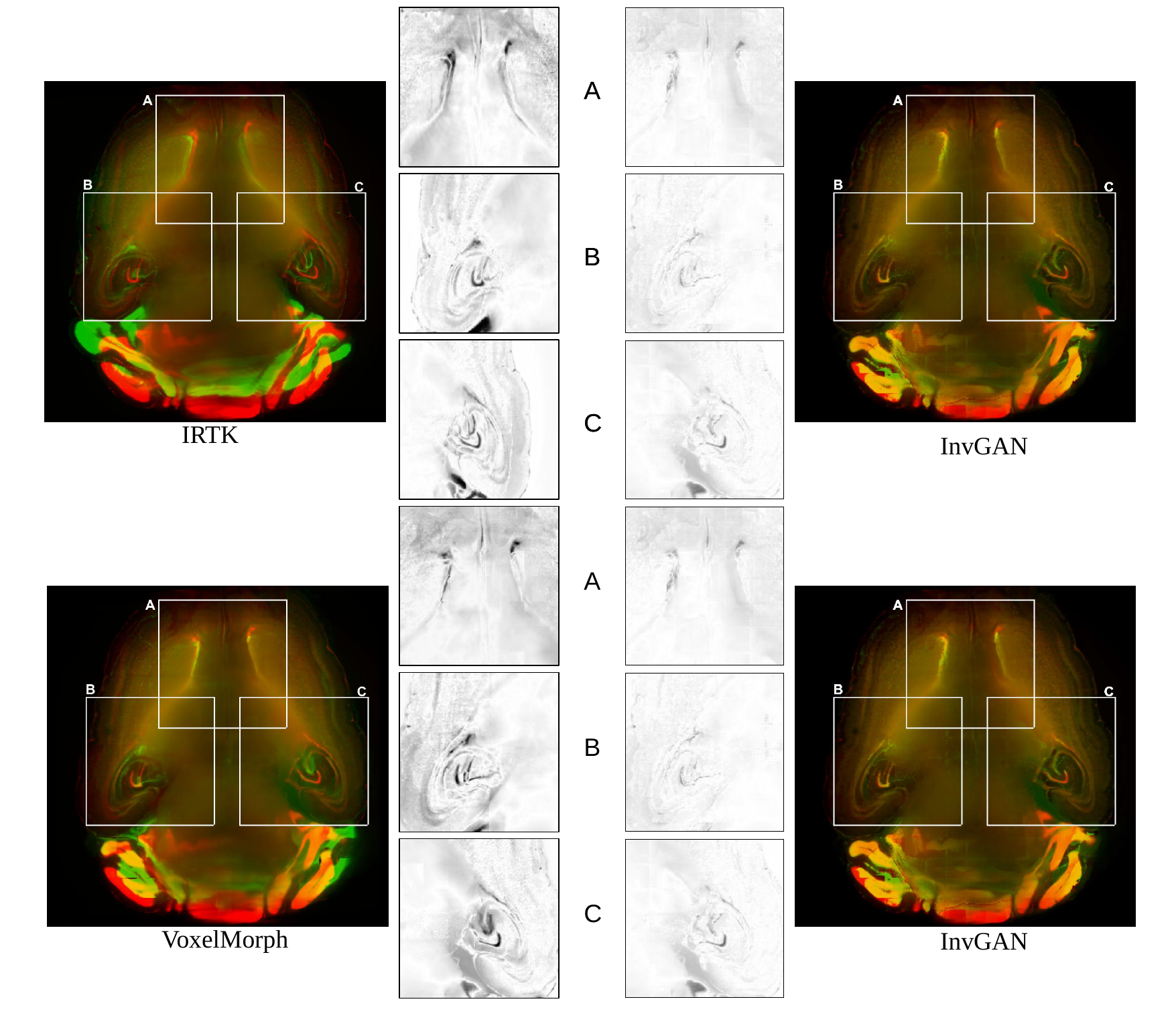}
    \caption{Visual Comparison at 25\% Resolution-2}
    \label{fig:25_compareDiff_2}
\end{figure*}

The qualitative comparison of the proposed method and other baseline methods are shown in Figure~\ref{fig:25_compareDiff_1} and \ref{fig:25_compareDiff_2}. We select the brain-1 registered by each each registration method and overlay on the target brain (brain-3). In Figures \ref{fig:25_compareDiff_1} and \ref{fig:25_compareDiff_2}, proposed InvGAN and other methods are compared side-by-side. The alignment in the registered images are visible in their color difference. When the registered image (green channel) perfectly aligns with reference image (red channel), perfect alignments are represented by yellow channel. Regions where alignments are not perfect, red and green channels are visible separately. To further extend the visualisation in more detail, we extract patches from three selected regions (Dentate Gyrus, left and right Hippocampal) from both registered image and reference image and calculate the difference between the patches. The difference image contains intensity values in the range +1 to -1 (since all brain volumes are intensity normalized from 0 to 1). For accurate alignment the intensity difference should be 0 in the difference image. For proper visualization of the difference image, we transform the difference image intensities into the range 0 to 255 using a linear triangular function centered at zero. The resulting transformed images thus contain intensity values 255 (0 in difference image) or white, in accurately aligned regions and 0 or dark in non-aligned regions.       
\\In Figures \ref{fig:25_compareDiff_1} and \ref{fig:25_compareDiff_2}, proposed InvGAN is compared with ANTS, Elastix,NiftyReg, IRTK and VoxelMorph. Between ANTS and InvGAN, it is difficult to differentiate which one is better based on overlay-ed image. But clear differences are found in the transformed difference images from three selected regions. In three regions, the number of dark spots in InvGAN registered brain is much smaller than the ANTS registered brain. This clearly indicates that InvGAN has better registration accuracy than ANTS.   In comparison to Elastix, the registration difference is clearly visible from the ovarlayed images. In Elastix registration, the red and green channels are not aligned perfectly and hence they are separate while in InvGAN registration they are aligned much accurately.The difference images further verifies the superior registration accuracy of InvGAN over Elastix. Similar pattern observed in case of NiftyReg and IRTK where alignment mismatches are clearly visible from the overlay images and further confirmed by local patches. The deep learning-based VoxelMorph on the other hand shows very similar pattern with ANTS. From the overlay image, registration performance of VoxelMorph is difficult to evaluate but the patches from three local regions clearly indicates that InvGAN beats VoxelMorph with large margin.              

\subsection{Results at 100\% Resolution}
The quantitative performances of InvGAN architecture and Elastix tool at 100\% resolution of the CUBIC data are presented in Table~\ref{Tab:100_results} and the qualitative results are shown in Figure~\ref{Fig:100registration}. Since no other conventional tools are able to register images to such extent, the results presented here only contain performance achieved by proposed InvGAN architecture and Elastix.

In Table~\ref{Tab:100_results}, the CC and MI scores are presented in different iterations of InvGAN training at 100\% resolution. The score before the registration is also provided to compare the registration performance. It takes a long time to improve the registration quality by InvGAN. At 10,000 iterations, the CC and MI scores improved slightly and started to degrade afterwards. For the CC score on brain-1, the elastix achieves the best scores with 0.8412 and 0.8676. For MI score, InvGAN is always higher than the elastix in brain-1. For brain-2, the CC score of InvGAN is smaller than Elastix while the MI score is slightly better.  
At 20,000 iterations, the performance becomes almost similar to the before-registration state for both brain. Slowly but gradually the accuracy continues to improve. 

At 88,000 iteration, InvGAN again improves. For brain-1, the CC score is reaches to 0.8076 and for brain-2 its reaches to 0.6278. In terms of mutual information, it achieves 0.9234 and 0.7801 respectively. At 100\% resolution, we trained and tested our method with only V-D side of CUBIC brains and the quantitative score represents that fact.     

The qualitative performances at 100\% resolution are presented in Figure~\ref{Fig:100registration} with different iterations. At each iteration, the color mapped overlay image shows the alignment quality and the gray-scale images present the image quality after registration. At 10,000 iterations, the alignment between brain-1 and brain-3 is not as expected but the image quality (shown in gray image) is good and has no box artifacts. In left and right hippocampal regions, the red and green channels of both brains are visible, which means the alignment in this region is not perfect. In the dentate-gyrus region, the differences between brains are still visible. At 60,000 iterations, with careful inspection it is found that the alignment in the hippocampal region improves slightly and it also improved in the dentate-gyrus region. At this iteration, the box artifacts start appearing. At iteration 80,000 and 88,000, the improvement of visual alignment between the brains is difficult to identify, while the image quality improves noticeably with no patch artifacts.

\begin{figure*}[!htb]
	\centering
	\begin{minipage}[t]{8cm}
		\includegraphics[width=\textwidth]{001_registered_88000_360.png}
		\subcaption{Brain-1 (InvGAN)}
	\end{minipage}%
	\begin{minipage}[t]{8cm}
		\includegraphics[width=\textwidth]{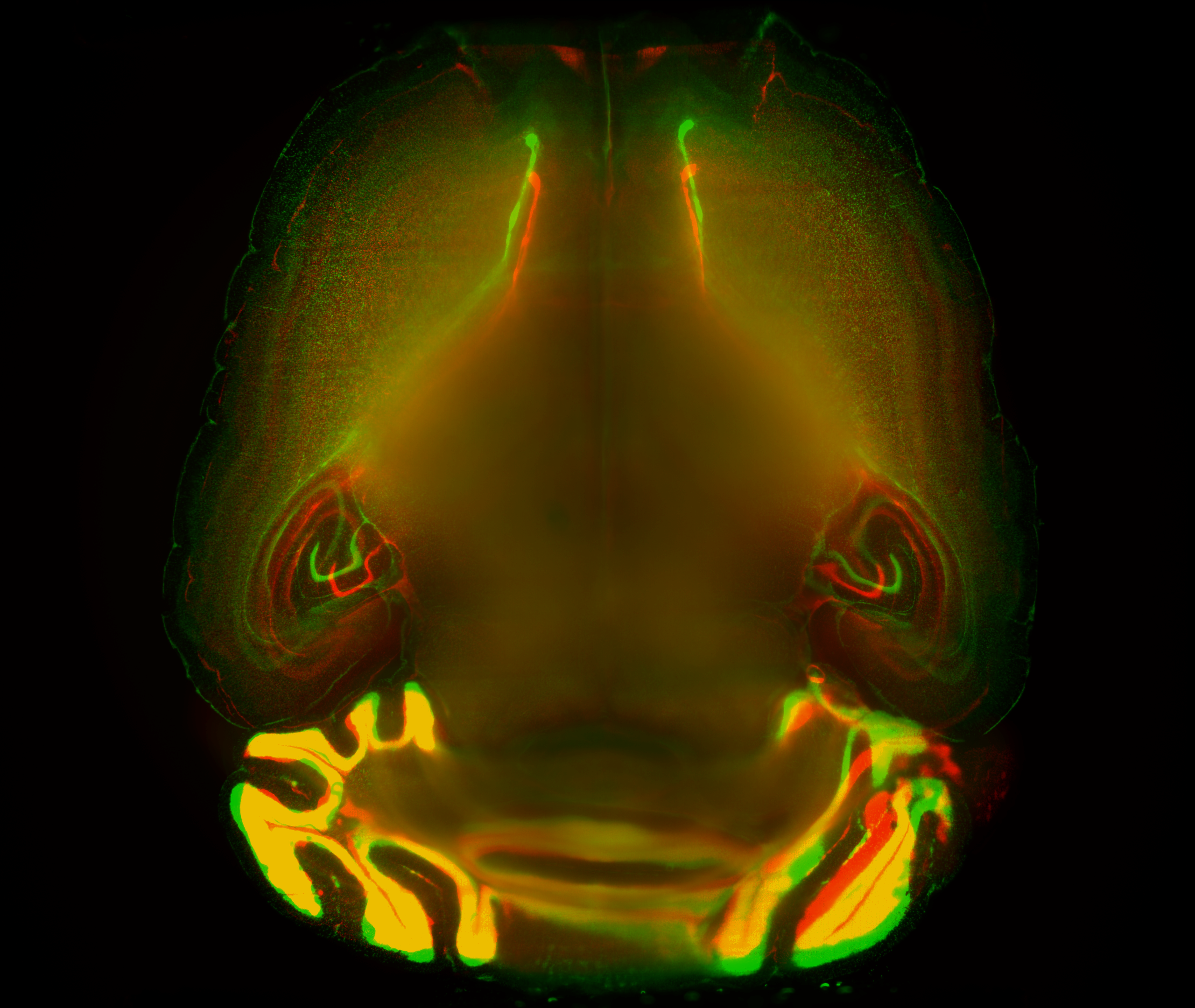}
		\subcaption{Brain-1 (Elastix)}
	\end{minipage}%
	\caption{Registration performance comparison at 100\% resolution}
	\label{Fig:100registration_cmp}
\end{figure*}

At 100\% resolution, the affine registration by ANTS tool takes around eight hours to align one pair. ANTS fails to apply deformable registration at 100\% resolution. The Elastix takes 27.5 hours of time to register one pair of 100\% resolution image. In contrary, proposed InvGan is extremely fast. It takes 6 mins to generate warp patches in the HPC environment and combining those patches in the big-data machine takes 2/3 mins per image, which takes in total 9 to 10 mins to register an image with 100\% resolution.
\\ To illustrate how our method is performing at 100\% in comparison to Elastix, Figure~\ref{Fig:100registration_cmp} is added. In Figure~\ref{Fig:100registration_cmp} the qualitative performance of two methods are compared side-by-side. The clear difference between 
reference image and Elastix registered image is visible in dentate-gyrus,left and right hippocampal regions. In InvGAN registered image, there are clear difference in left hippocampal but in right hippocampal the red and green channels are almost aligned. Since we trained our model with affine aligned V-D side only, the misalignment on the other part of the brain is expected.    
In Dentate Gyrus, the Elastix completely fails to align while our method aligns two brain more perfectly.The high quantitative score of Elastix is due to the fact that it aligns the Cerebellum region, therefore, the lower part of the brains more accurately than the dentate-gyrus and both hippocampal regions. The alignment in this region does not provide any reasonable conclusion since this region extensively varies from brain to brain.

The qualitative and quantitative results at 100\% resolution indicates that the proposed method is applicable to very high resolution images. Considering the fact that only V-D side is used for training, its consistent performance  in dentete gyrus and hippocampal regions compared to Elastix clearly shows its potential for very high resolution image registration.     

\begin{table}[!htb]
\begin{center}
	\caption{\small  Performance at 100\% Resolution}
	\label{Tab:100_results}
	\small
	\begin{tabular}{l {c}{c} {c} {c}}
		\hline
		Methods & brain-1 &brain-1 & brain-2 & brain-2\\
		        &  CC        &MI  &CC &MI\\
		\hline
		BEFORE      		            &0.7707  	&0.9640     &0.5872  	&0.8049\\
		Elastix                &\textbf{0.8412}     &0.8018     &\textbf{0.8676}     &\textbf{0.8385}\\
		InvGAN 10000					&0.8046		&\textbf{1.0026}		&0.6307		&0.8378\\
		InvGAN 20000					&0.7776		&0.9195		&0.5949		&0.7689\\
		InvGAN 40000					&0.7837		&0.9189		&0.6171		&0.7859\\
		InvGAN 60000					&0.7944		&0.9604		&0.6568		&0.8315\\
		InvGAN 80000					&0.8041		&0.9579		&0.6298		&0.8250\\
		InvGAN 88000					&0.8076		&0.9234		&0.6278		&0.7801\\
		\hline
	\end{tabular}
\end{center}
\end{table}

\begin{figure*}[!htb]
	\begin{center}
		\begin{minipage}[t]{8cm}
			\includegraphics[width=\textwidth]{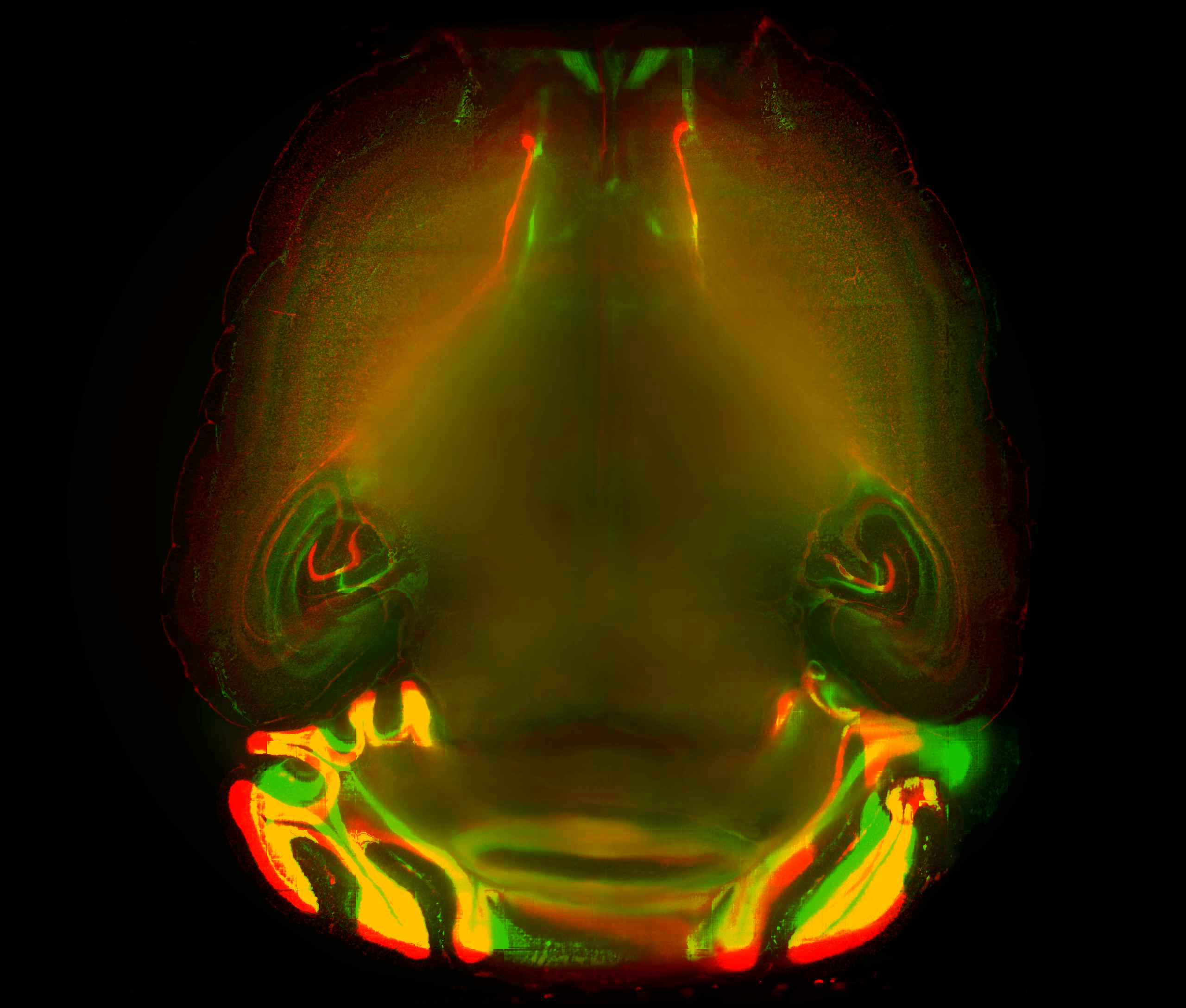}
			\subcaption{Brain-1 Overlay at 10,000 iteration}
		\end{minipage}%
		\begin{minipage}[t]{8cm}
			\includegraphics[width=\textwidth]{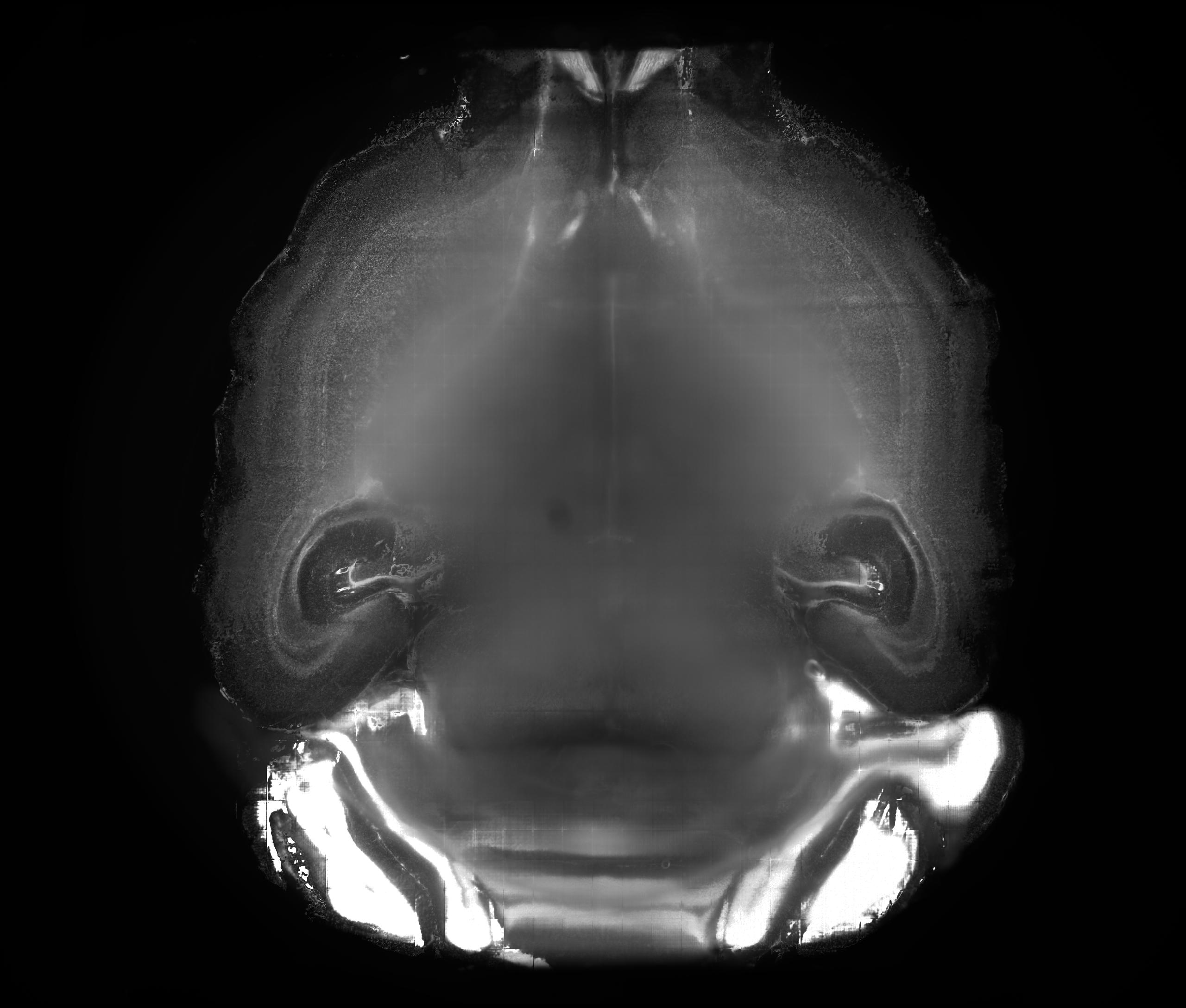}
			\subcaption{Brain-1 at 10,000 iteration}
		\end{minipage}%
		\vfill
		\begin{minipage}[t]{8cm}
			\includegraphics[width=\textwidth]{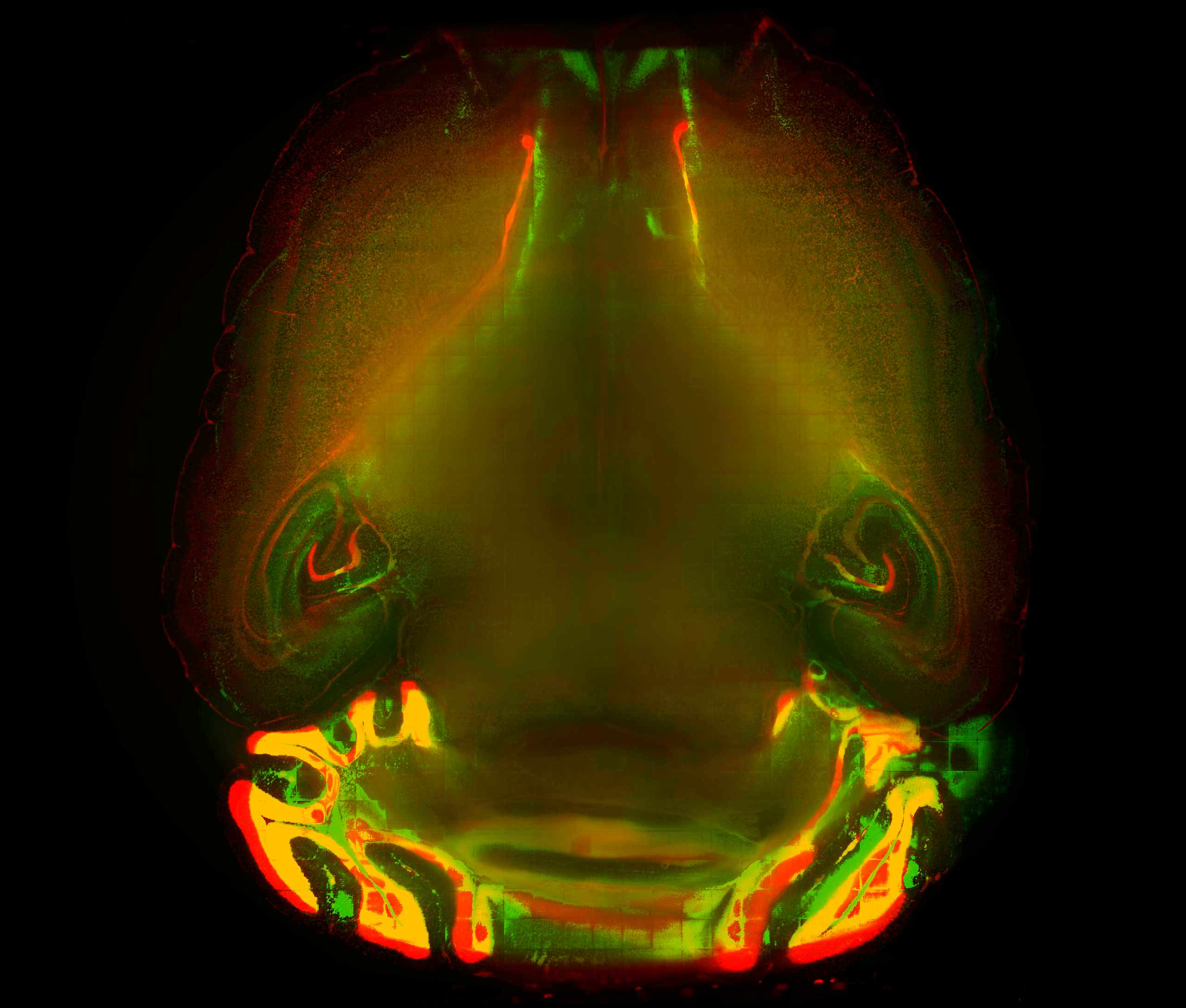}
			\subcaption{Brain-1 Overlay at 60,000 iteration}
		\end{minipage}%
		\begin{minipage}[t]{8cm}
			\includegraphics[width=\textwidth]{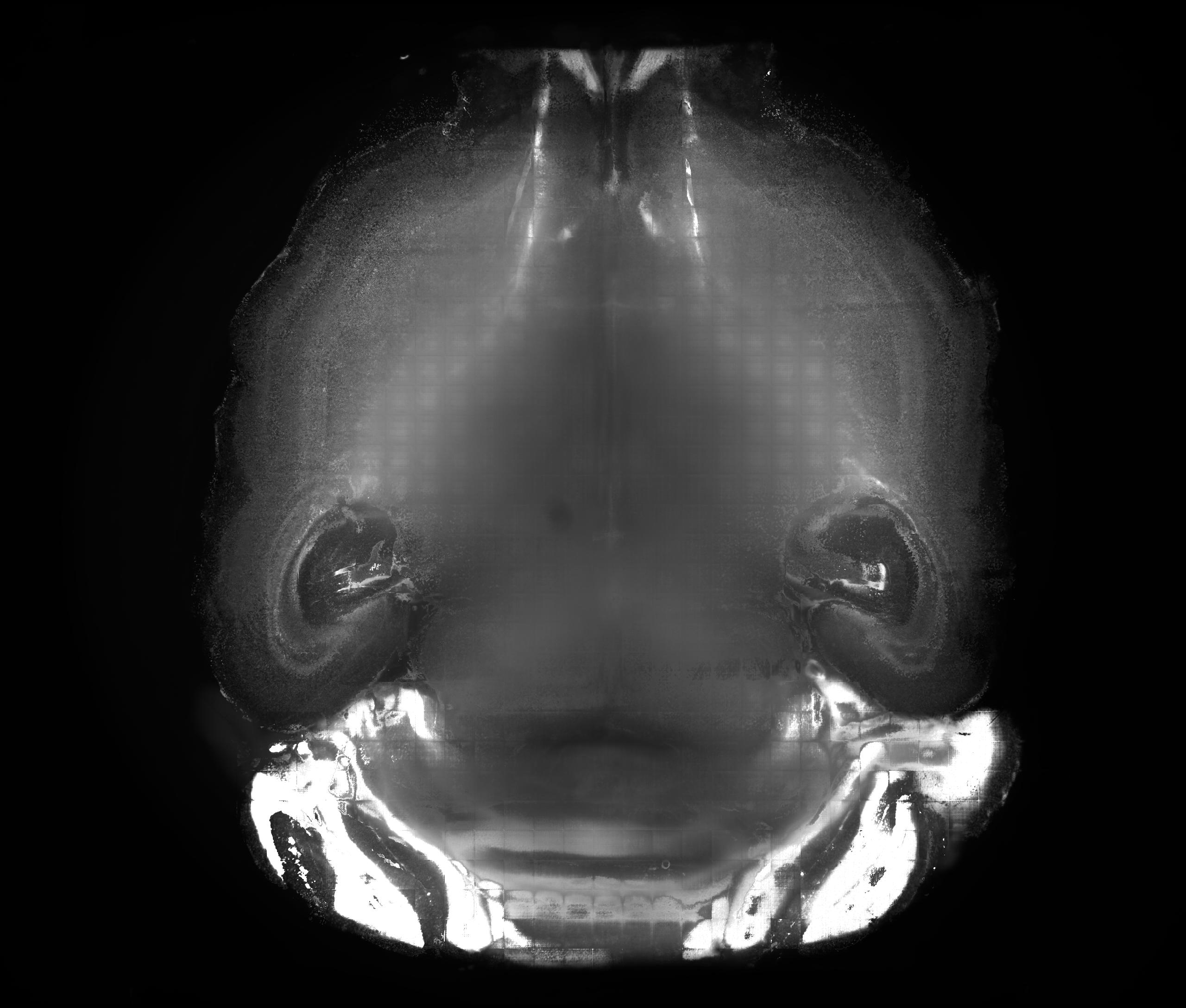}
			\subcaption{Brain-1 at 60,000 iteration}
		\end{minipage}%
		\vfill
		\begin{minipage}[t]{8cm}
			\includegraphics[width=\textwidth]{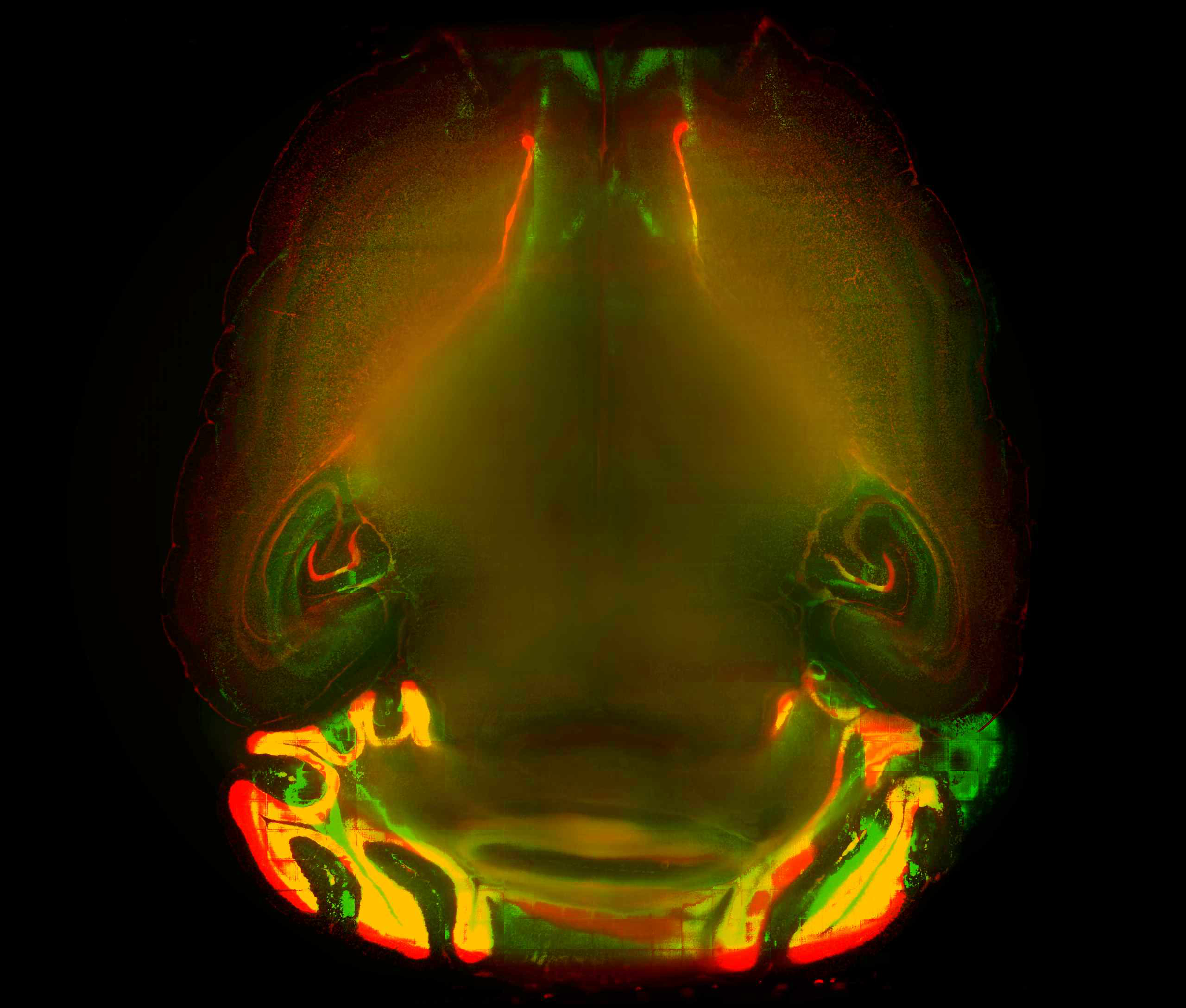}
			\subcaption{Brain-1 Overlay at 80,000 iteration}
		\end{minipage}%
		\begin{minipage}[t]{8cm}
			\includegraphics[width=\textwidth]{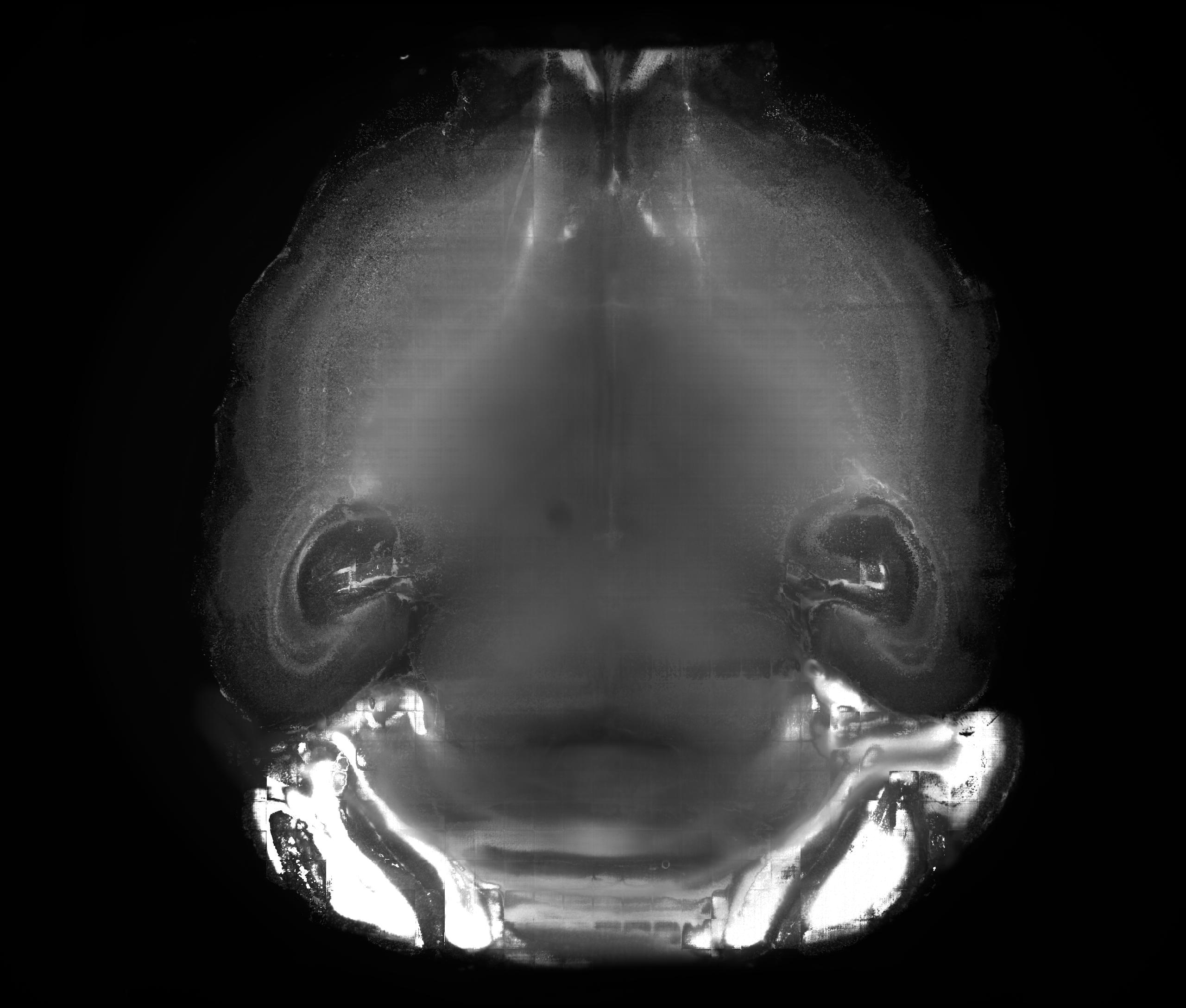}
			\subcaption{Brain-1 at 80,000 iteration}
		\end{minipage}%
	\end{center}
\end{figure*}
\begin{figure*}
	\ContinuedFloat
	\centering
	\begin{minipage}[t]{8cm}
		\includegraphics[width=\textwidth]{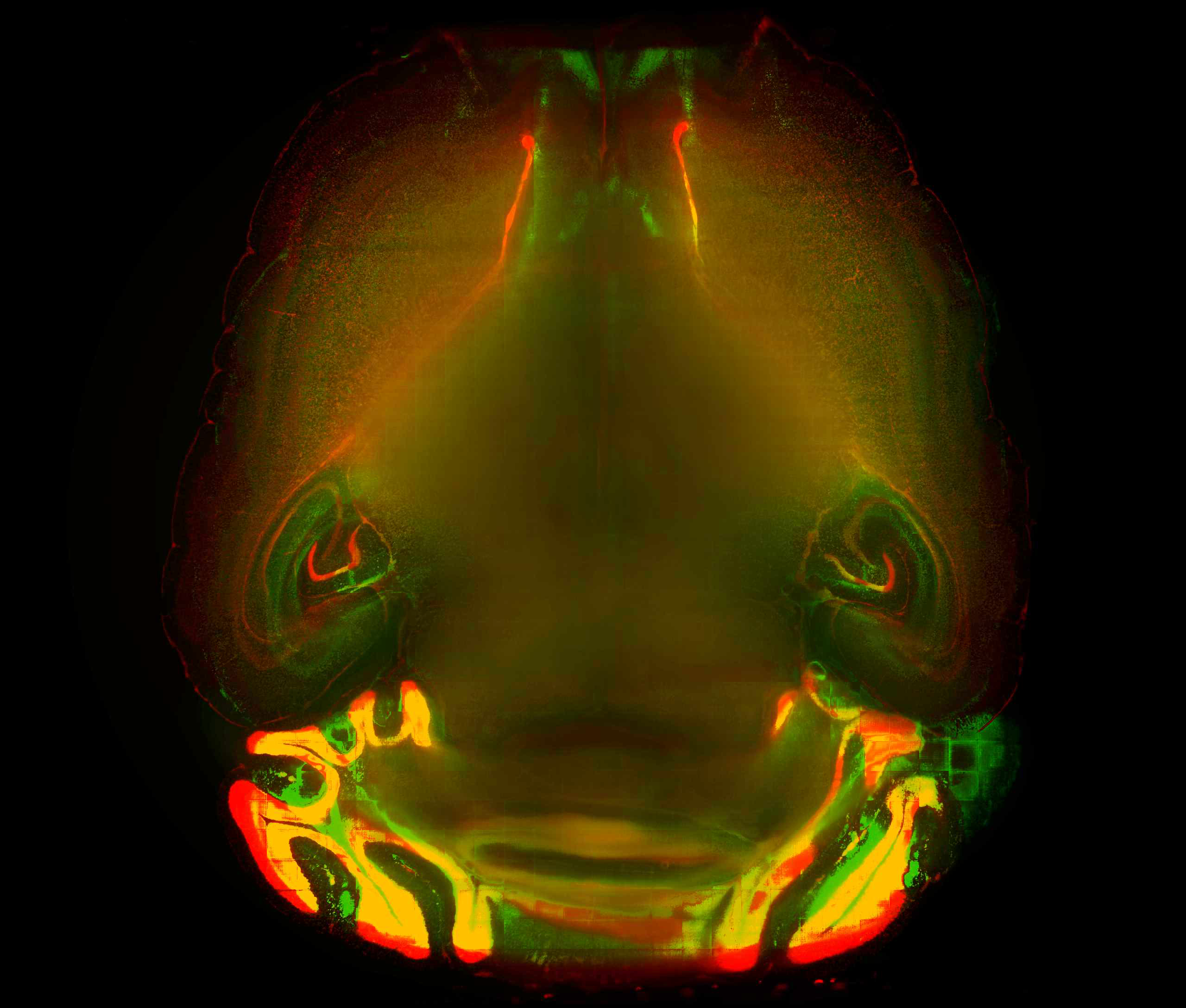}
		\subcaption{Brain-1 Overlay at 88,000 iteration}
	\end{minipage}%
	\begin{minipage}[t]{8cm}
		\includegraphics[width=\textwidth]{001_registered_88000gray_360.jpg}
		\subcaption{Brain-1 at 88,000 iteration}
	\end{minipage}%
	\caption{Registration performance at 100\% Resolution with Different Iteration}
	\label{Fig:100registration}
\end{figure*}

\subsection{Landmark Validation}
\begin{figure*}
	\begin{center}
		\begin{minipage}[t]{4cm}
			\includegraphics[width=\linewidth,height=4cm]{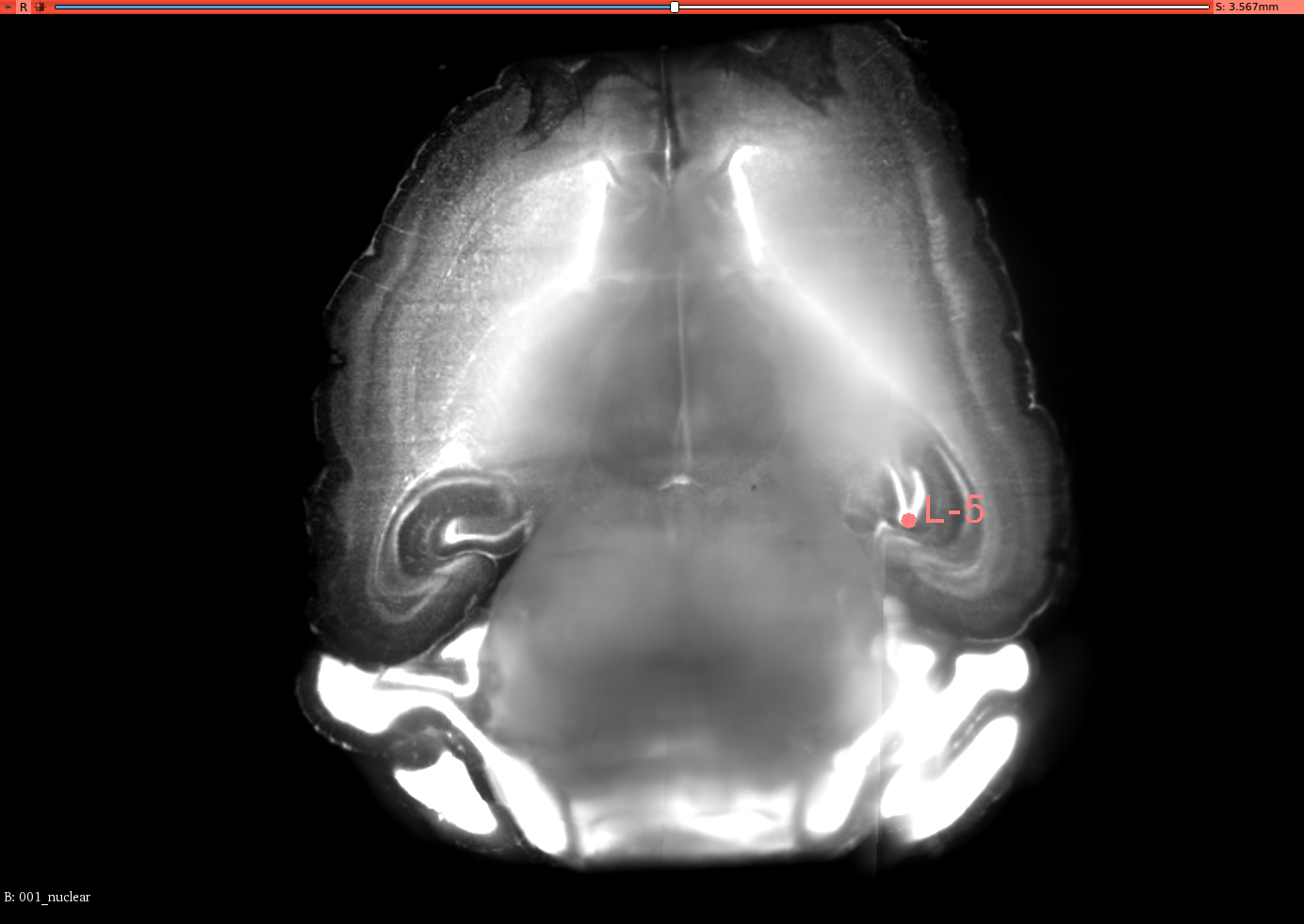}
			\subcaption{Brain-1}
		\end{minipage}%
		\begin{minipage}[t]{4cm}
			\includegraphics[width=\linewidth,height=4cm]{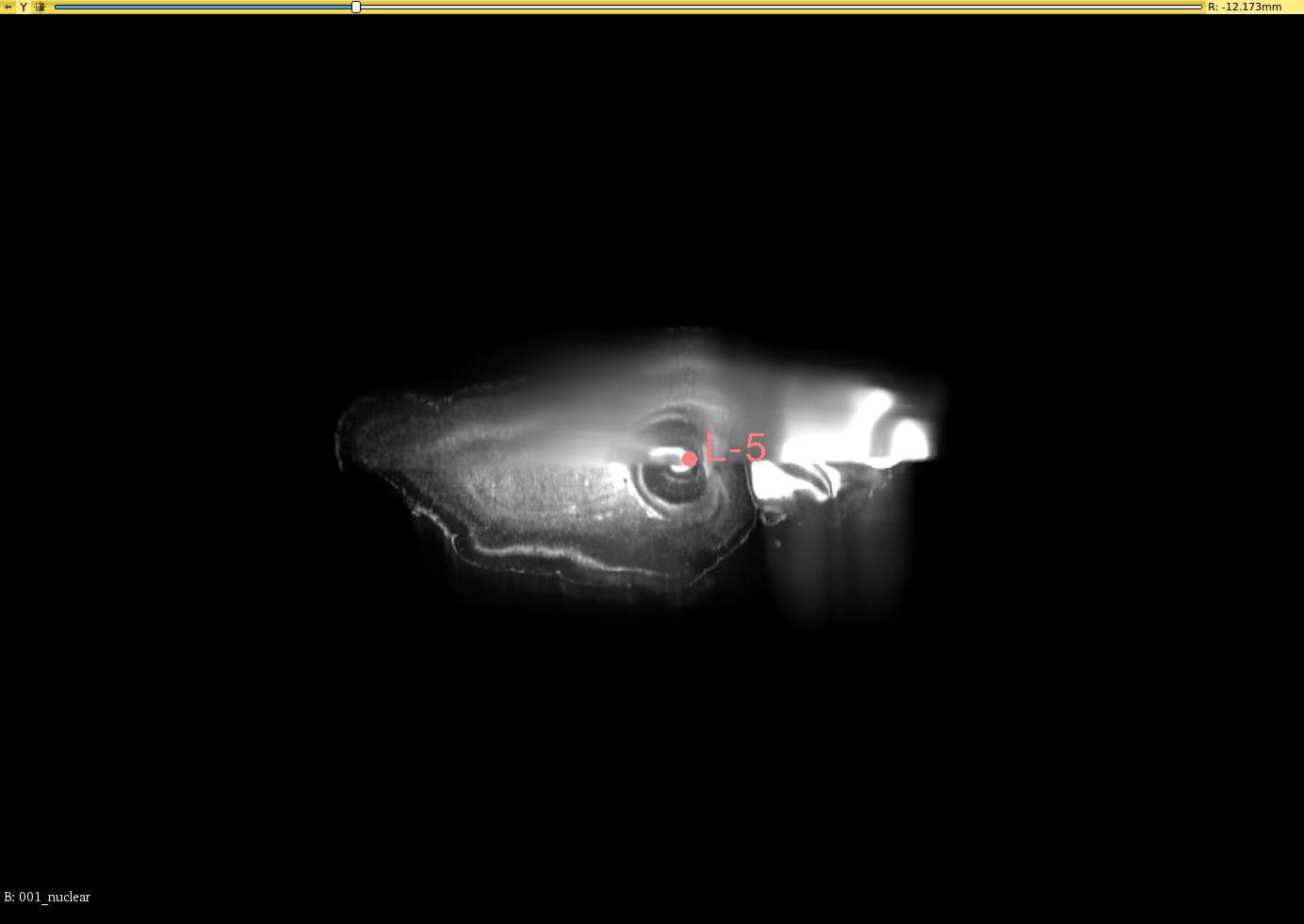}
			\subcaption{Segittal View}
		\end{minipage}%
		\begin{minipage}[t]{4cm}
			\includegraphics[width=\linewidth,height=4cm]{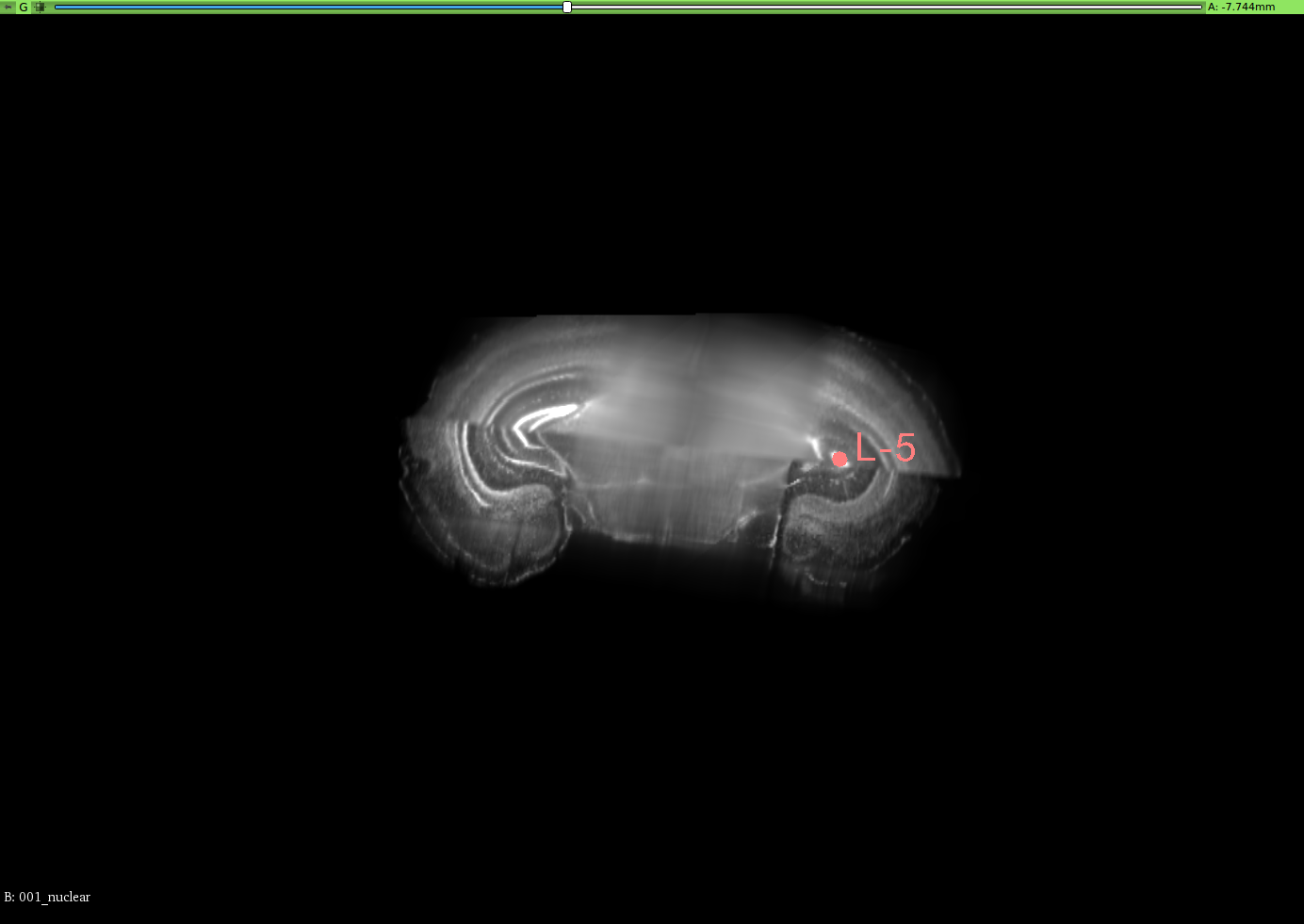}
			\subcaption{Coronal View}
		\end{minipage}%
		\vfill
		\begin{minipage}[t]{4cm}
			\includegraphics[width=\linewidth,height=4cm]{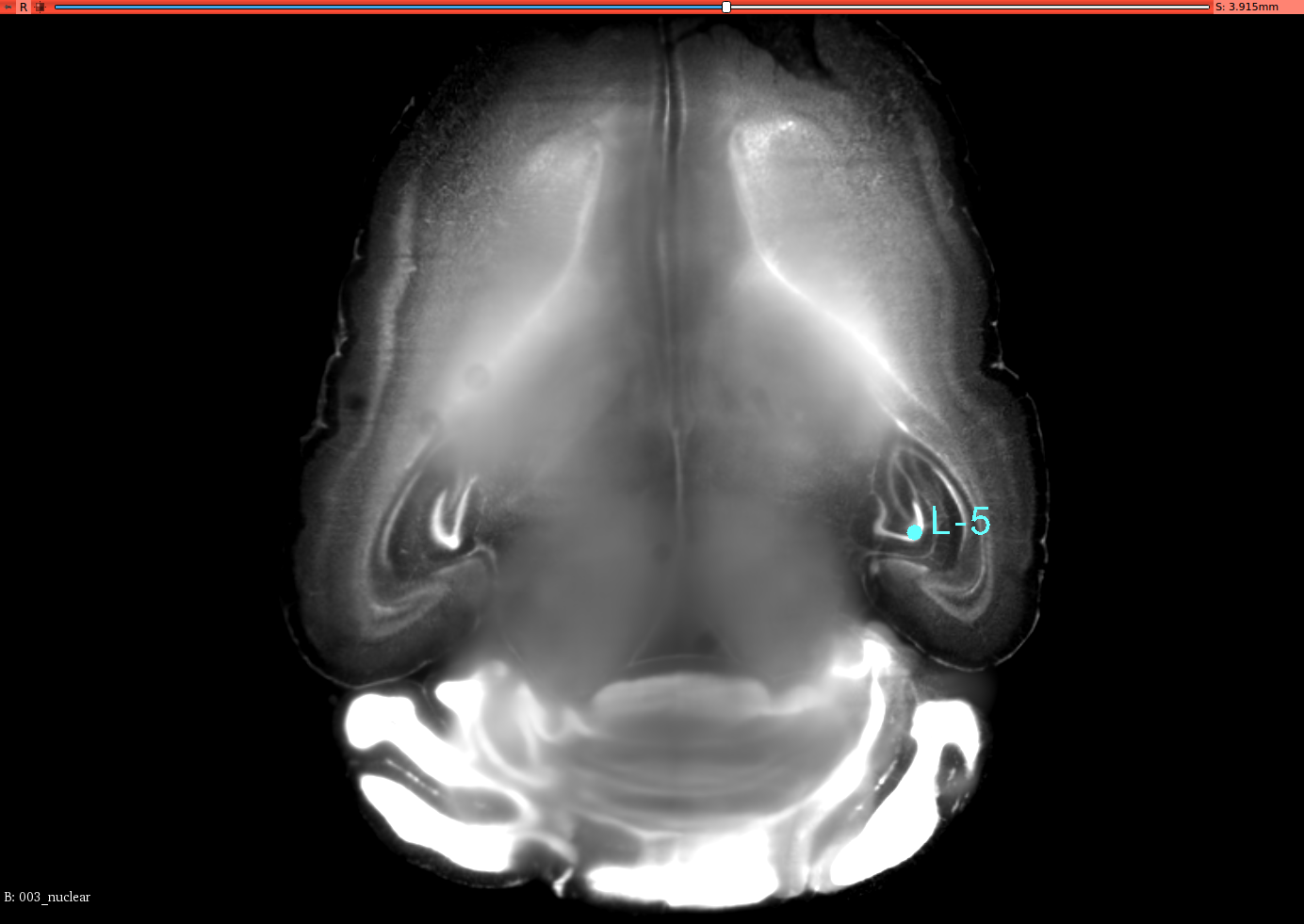}
			\subcaption{Corresponding Points in Brain003}
		\end{minipage}%
		\begin{minipage}[t]{4cm}
			\includegraphics[width=\linewidth,height=4cm]{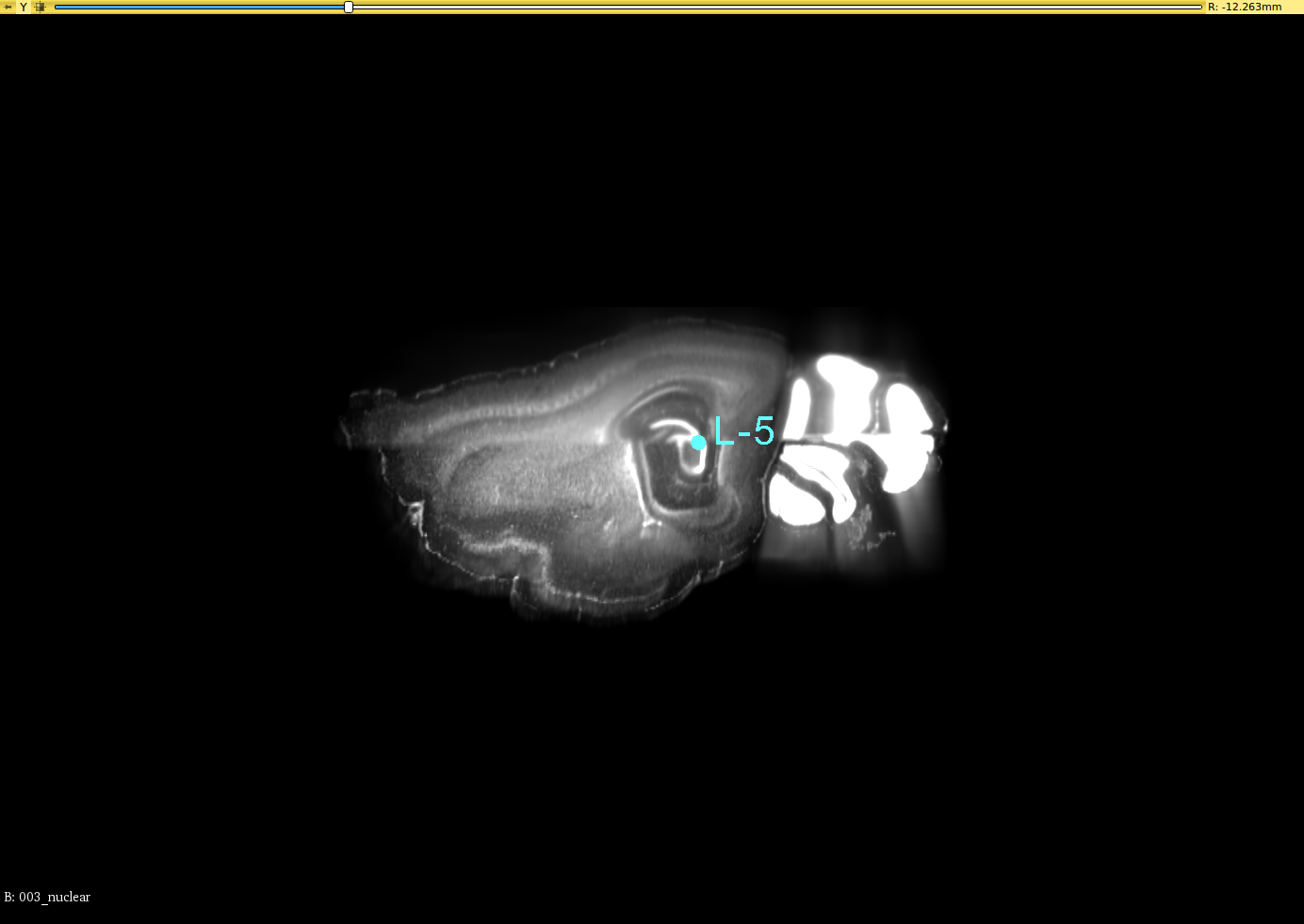}
			\subcaption{Segittal View}
		\end{minipage}%
		\begin{minipage}[t]{4cm}
			\includegraphics[width=\linewidth,height=4cm]{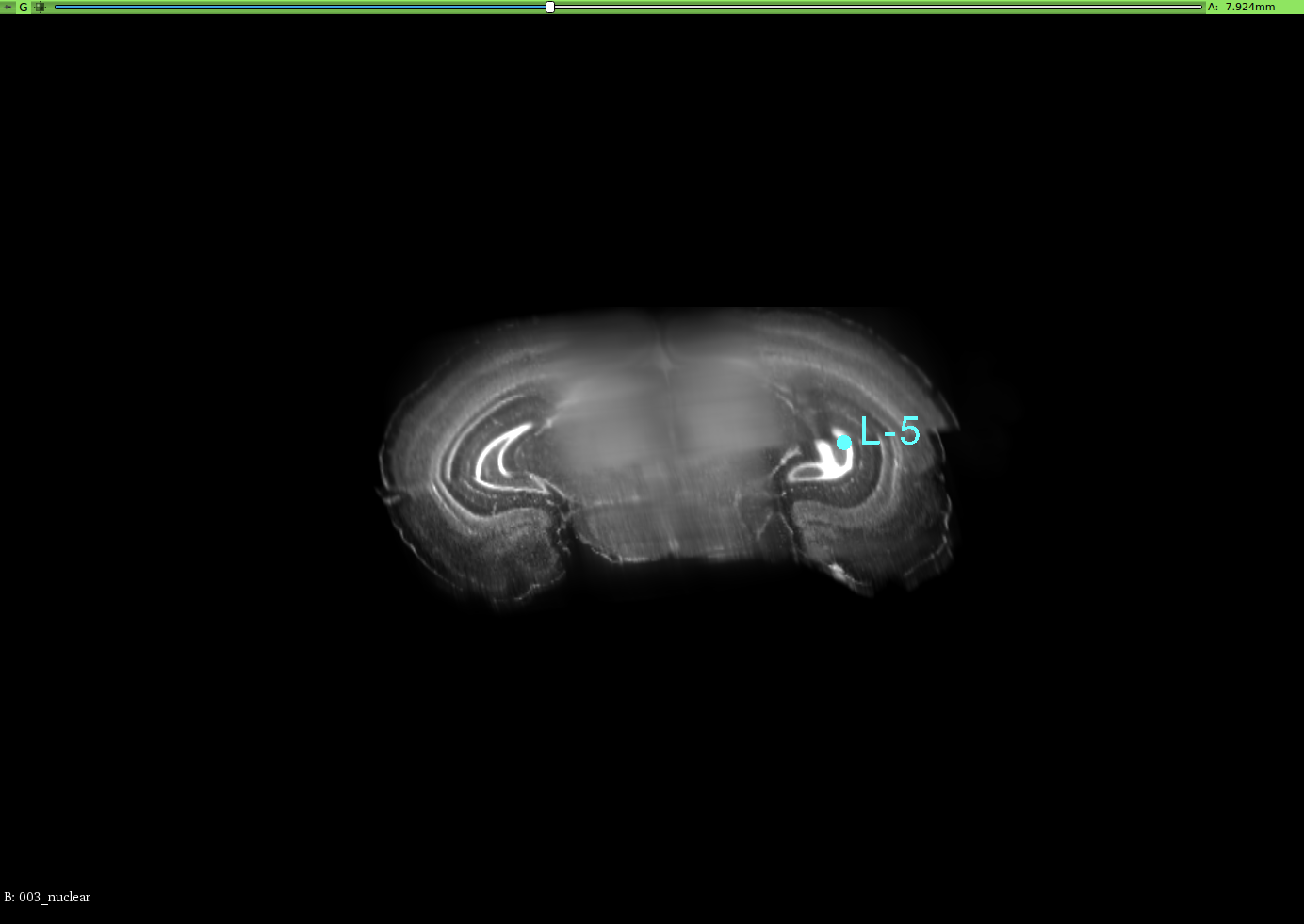}
			\subcaption{Coronal View}
		\end{minipage}%
		\vfill
		\begin{minipage}[t]{4cm}
			\includegraphics[width=\linewidth,height=4cm]{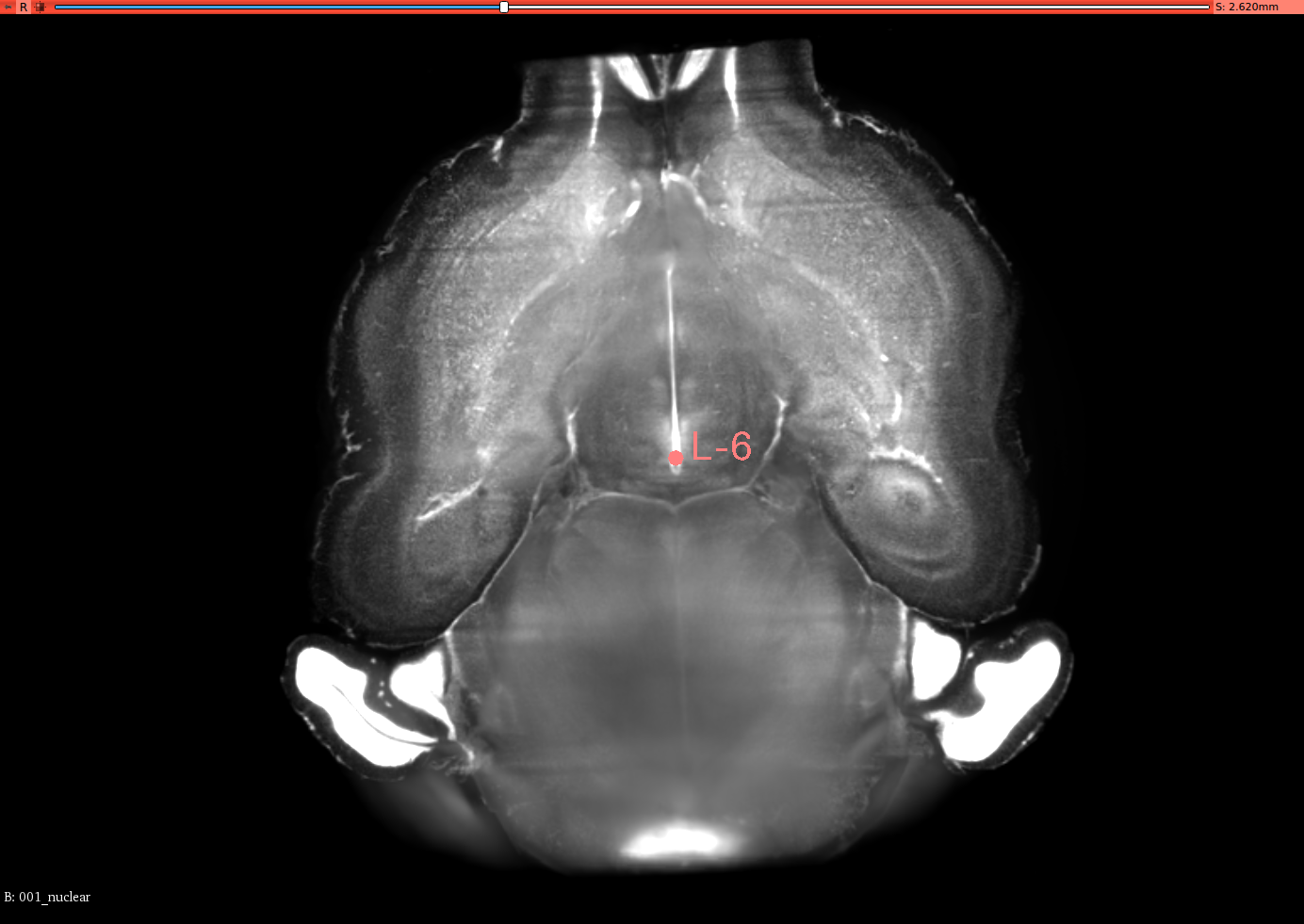}
			\subcaption{Brain-2}
		\end{minipage}%
		\begin{minipage}[t]{4cm}
			\includegraphics[width=\linewidth,height=4cm]{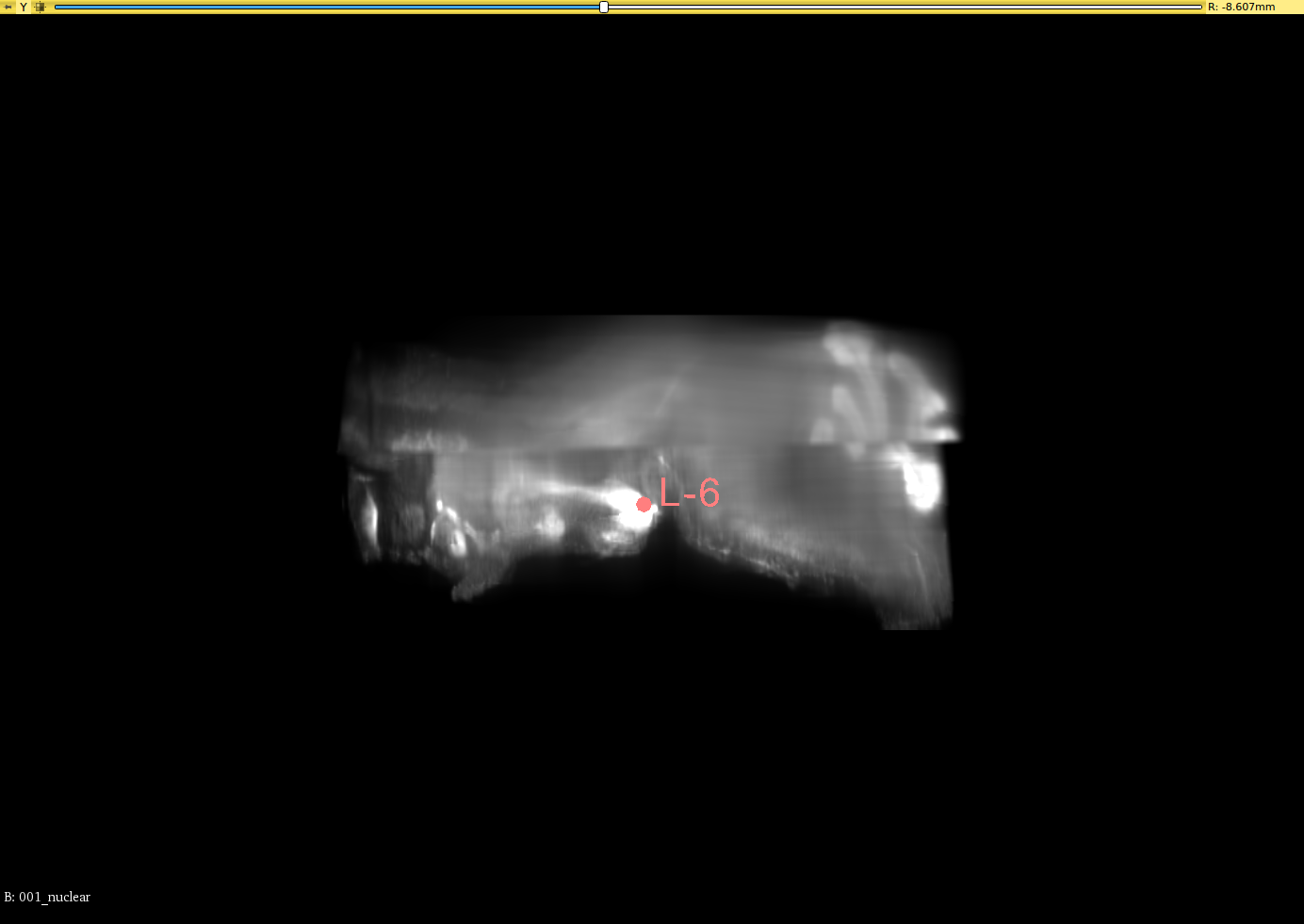}
			\subcaption{Segittal View}
		\end{minipage}%
		\begin{minipage}[t]{4cm}
			\includegraphics[width=\linewidth,height=4cm]{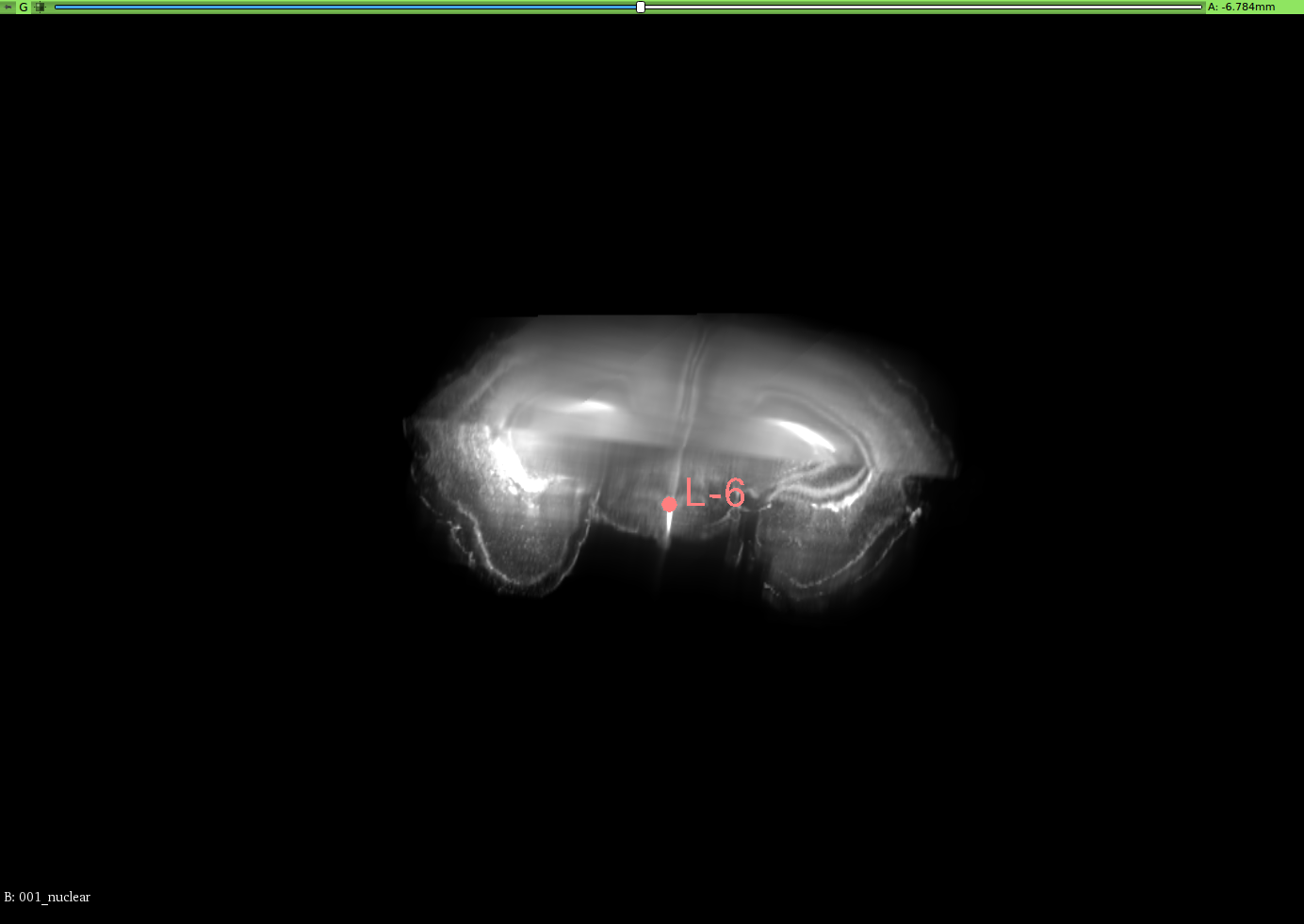}
			\subcaption{Coronal View}
		\end{minipage}%
		\vfill
		\begin{minipage}[t]{4cm}
			\includegraphics[width=\linewidth,height=4cm]{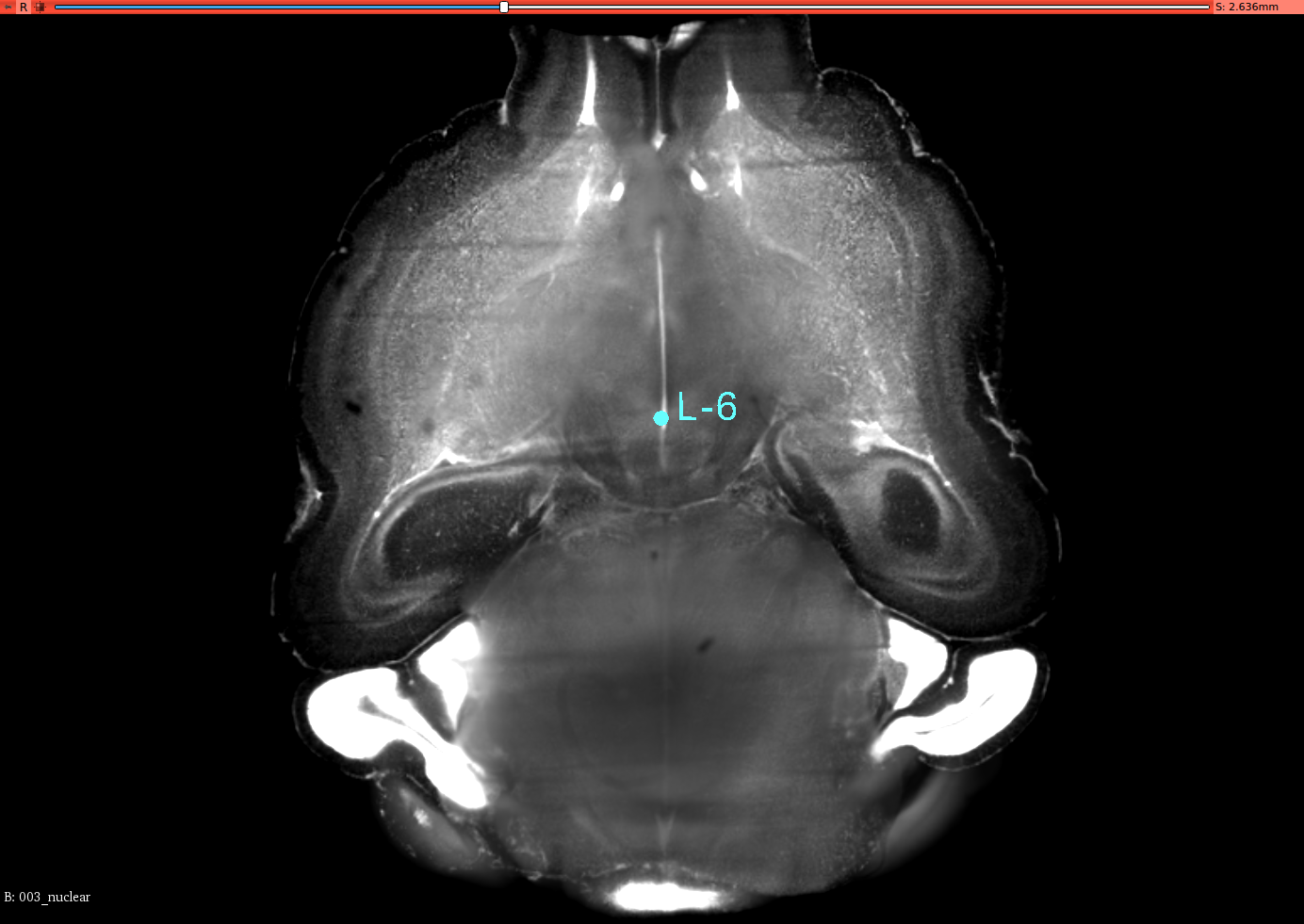}
			\subcaption{Corresponding Points in Brain-3}
		\end{minipage}%
		\begin{minipage}[t]{4cm}
			\includegraphics[width=\linewidth,height=4cm]{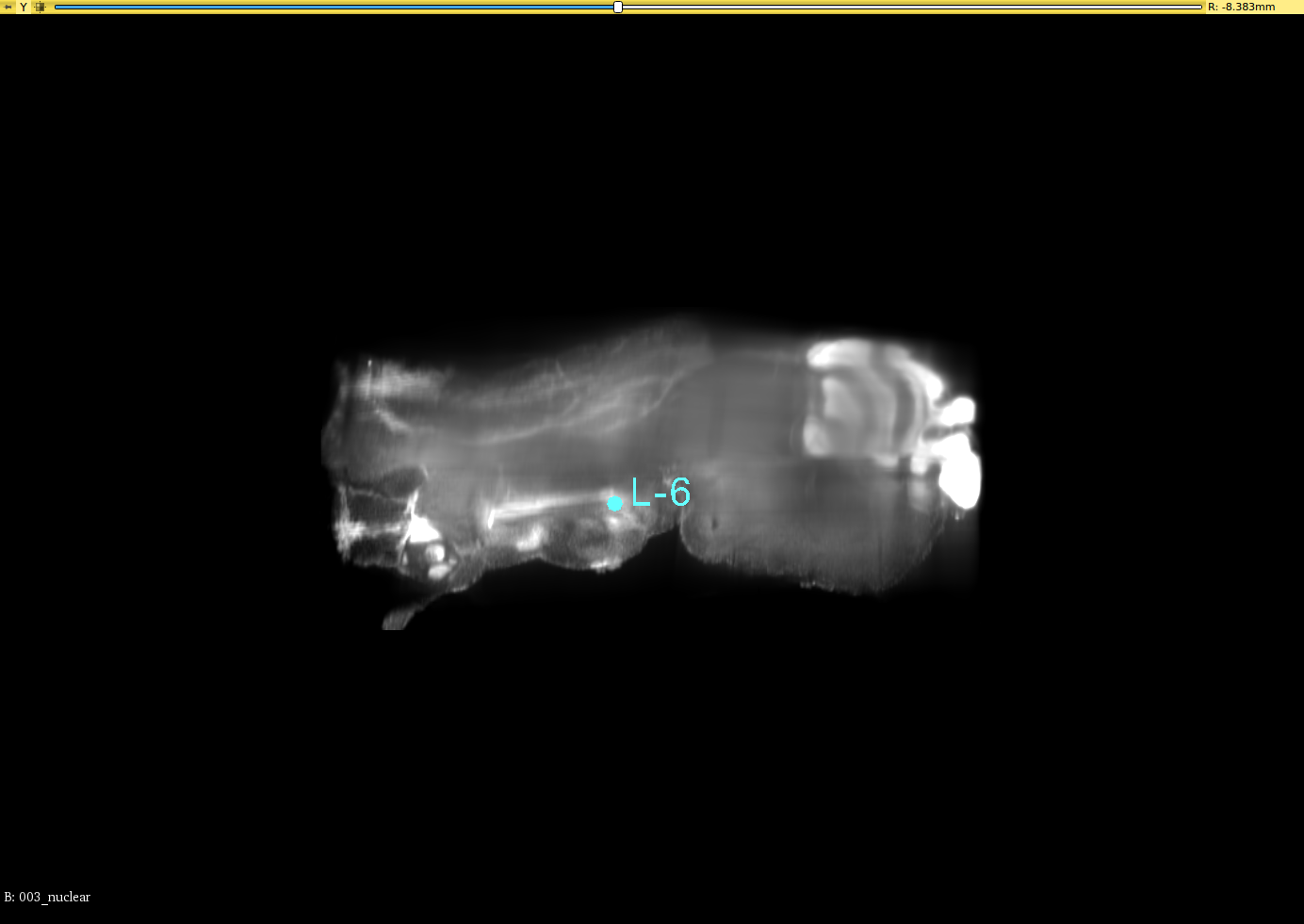}
			\subcaption{Segittal View}
		\end{minipage}%
		\begin{minipage}[t]{4cm}
			\includegraphics[width=\linewidth,height=4cm]{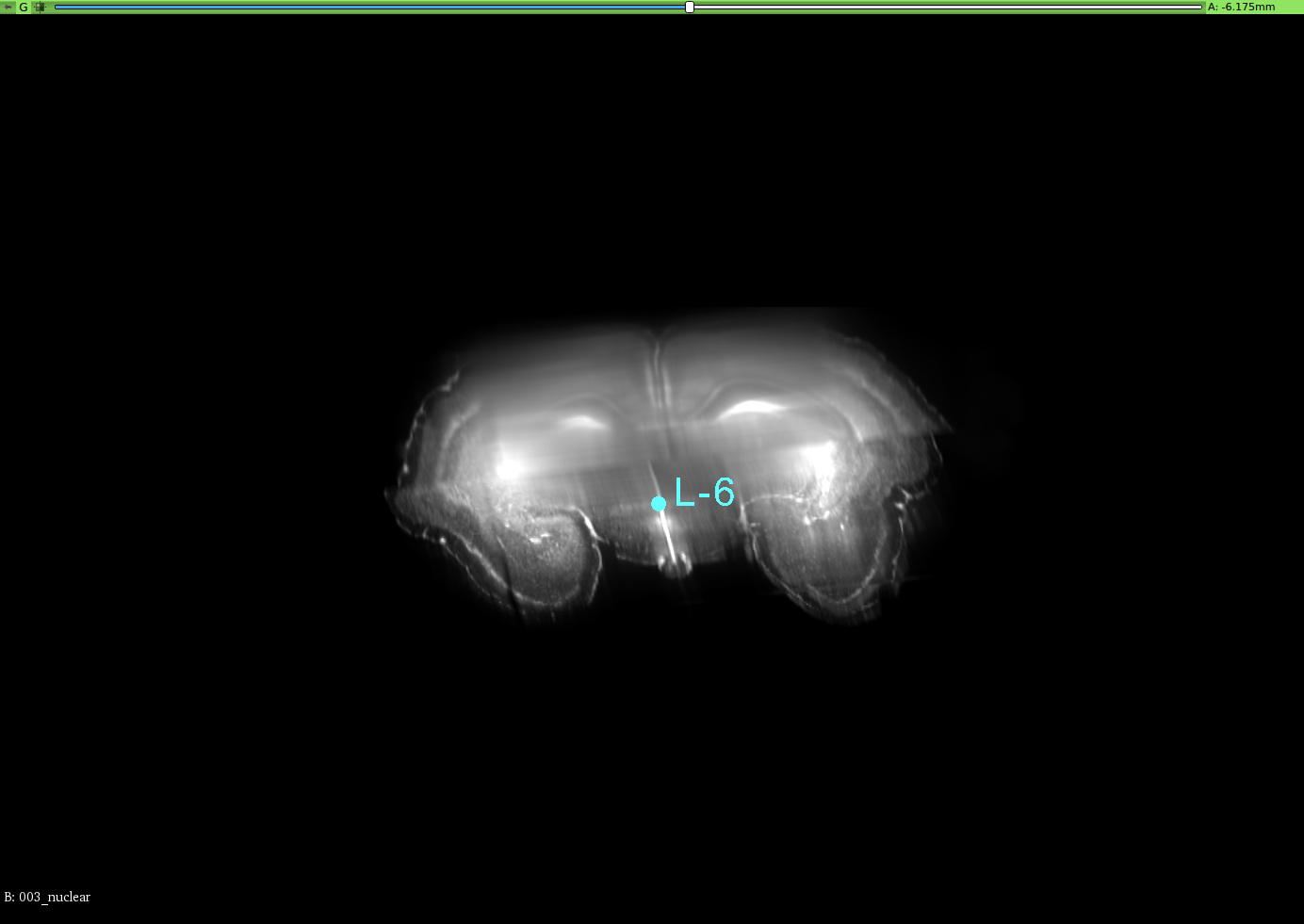}
			\subcaption{Coronal View}
		\end{minipage}%
		\vfill
		\caption{3D Landmarks}
		\label{ch7:Fig:Landmarkrs_set2}
	\end{center}
\end{figure*}

To validate the registration performance of the proposed methods and comparing the baseline methods in a more objective manner, a landmark registration test is conducted. In the CUBIC dataset, three brains are used to test the registration performance. In the landmark test, the same dataset is used for the performance validation. 12 landmarks are selected and all of these landmarks are selected where their positions vary in all three axis. 3D slicer tool is used to select the landmarks for this experiment from the CUBIC brains. A set of selected landmarks are shown in Figure~\ref{ch7:Fig:Landmarkrs_set2}.
Table~\ref{Tab:Landmark} shows the results of the 3D landmark registration by proposed InvGAN method and other baseline methods. The Euclidean distance between the registered landmarks and reference landmarks are presented in mm. For optimisation-based ANTS and Elastix tools, the same parameter sets are used to register landmarks selected from moving image and fixed image. After the registration, the output point locations are compared with reference point locations in the fixed image. For the deep-learning-based VoxelMorph and InvGAN, the deformation values in X,Y and Z are extracted from the same voxel location of the selected landmark’s voxel location. After applying deformation to the landmarks, the new position is compared with that of corresponding reference points in the fixed image.

In Table~\ref{Tab:Landmark}, distances between fixed image landmarks and registered landmarks are presented in columns. In this test, NiftyReg and IRTK both are excluded due to the lack of technical documentation provided for landmark registration.
In Table~\ref{Tab:Landmark}  brainwise scores performed by each registration tool and the average of all 12 landmarks as well as their standard deviation are presented. The smaller the scores, the better the registration performance is.
For brain-1 proposed InvGAN achieves least average distance with 0.1547mm. The VoxelMorph achieves second best score with 0.3362mm while ANTS become third in position with 0.3685mm. Elastix become the lest performer with 4.4555mm. In brain-2, ANTS achieves the best result with 0.1938mm while proposed InvGAN achieves 0.2926mm. The VoxelMorph improves its performance with 0.2693mm. The Elastix tool remains again the least performer in brain-2 and its performance in brain-2 is even worse than brain-1 with more than 6mm average distance.    

\begin{table}[ht]
	\begin{center}
		\caption{\small Landmark Test for 3D Landmarks}
		\label{Tab:Landmark}
		\small
		\begin{tabular}{l p{10mm} p{10mm} |p{10mm} p{10mm}}
			\hline
			Landmarks & brain 1(l2) &brain 2(l2) & brain 1(l2) &brain 2(l2)\\
			\hline
			&\multicolumn{2}{c}{\textbf{ANTS}} 	&\multicolumn{2}{c}{\textbf{InvGAN}} \\
			\hline
			Point 1     &0.1915		&0.1549          &0.1778	&0.2349\\
			Point 2	 	&0.2706		&0.0876			 &0.1115	&0.2735\\
			Point 3     &0.2509		&0.2392		     &0.3448	&0.3573\\
			Point 4		&0.1742		&0.2102		     &0.4336	&0.1890\\
			Point 5		&0.3452		&0.1386		     &0.2008	&0.3894\\
			Point 6	 	&0.4177		&0.2362		     &0.0315	&0.2898\\
			Point 7	 	&0.8581		&0.2106		     &0.0545	&0.2482\\
			Point 8	 	&0.2418		&0.1402		     &0.0652	&0.2090\\
			Point 9 	&0.6462		&0.4425 	     &0.1676	&0.4907\\
			Point 10	&0.3461		&0.0934 	     &0.0815	&0.3050\\
			Point 11	&0.4529		&0.0703 	     &0.0552	&0.3017\\
			Point 12 	&0.1648		&0.1403			 &0.0170 	&0.1841\\   
			\hline
			Avg 		&0.3685 	&\textbf{0.1938} 	   &\textbf{0.1547}	&0.2926\\
			\hline 
			Std			&0.1924		&0.1035			&0.1238	&0.0830\\
			\hline
			&\multicolumn{2}{c}{\textbf{Elastix}} &	\multicolumn{2}{c}{\textbf{VoxelMorph}}\\
			\hline
			Point 1     &2.8008		&7.7366  	     &0.3011	    &0.1765\\ 
			Point 2     &1.9376		&3.8387			 &0.1330 	&0.2711\\	
			Point 3	 	&2.4545		&7.3544		     &0.4846 	&0.3547 \\
			Point 4     &2.0241		&7.6805		     &0.3240 	&0.1844\\
			Point 5	 	&7.3584		&7.1101		     &0.1184	    &0.2565\\
			Point 6	 	&8.0046		&7.4711		     &0.3194	    &0.2782\\
			Point 7	 	&5.2184		&7.3308		     &0.6680 	    &0.1745\\
			Point 8	 	&6.3149		&7.4293		     &0.1770 	    &0.2236\\
			Point 9	 	&6.3055		&7.0961		     &0.5091 	    &0.4889\\
			Point 10 	&3.9807		&7.2472		     &0.4329 	    &0.3494\\
			Point 11	&3.3438		&2.4100		     &0.4260	    &0.2094\\
			Point 12	&3.7224		&0.4873			 &0.2982 	    &0.1304\\
			\hline
			Avg 		&4.4555 			&6.0994 	&0.3362	    &0.2693	   \\
			\hline 
			Std			&2.0305				&2.3358		&0.1572	    &0.0999	  	\\
			\hline
		\end{tabular}
	\end{center}
\end{table}

\subsection{Computation Time}
The side-by-side comparisons of computation time at 25\% and 100\% resolutions are presented in Table\ref{Table:computationtime}. The computation time plays as a significant indicator of performance, efficiency and applicability of the comparing methods on tissue cleared data. In Table\ref{Table:computationtime}, the registration time at 25\% of all the traditional tools takes significantly longer time for both affine and deformable registration than the deep-learning based tools.
Proposed InGAN and its deep-learning based counterpart VoxelMorph take approximately 1 min for deformable registration. The VoxelMorph takes only 55s which is slightly better than proposed InvGAN but with the cost of registration accuracy (\textit{see Table\ref{Tab:25_quantitative}}). 
The performance difference more significantly evident when the resolution increases. Most of the traditional tools fails to register at 100\% resolution. The ANTS only performs affine registration and fails for the deformable.For affine only, it takes around 8 hours of time for a single pair registration. Elastix, on the other hand performs both affine and deformable registration with the exponential increase in registration time. Registration time for both deep learning based methods are significantly smaller than their traditional counterparts at 100\%. A similar pattern in registration time at 100\% resolution is shown by both deep-learning based methods. Like 25\%, at 100\% the proposed InvGAN takes slightly longer time (\textit{approximately 10 mins}) than VoxelMorph (\textit{approximately 7 mins}). By considering the performance measures in Table~\ref{Tab:100_results} and landmark results in Table~\ref{Tab:Landmark} it is evident that the fast registration time of VoxelMorph is achieved with cost of registration accuracy which is not the case for proposed InvGAN network. Considering high registration accuracy of InvGAN, its slight longer registration time is acceptable for analysis pipeline like CUBIC.                
\begin{table}[h!]

	\small
	\begin{center}
		\caption{Registration Time at 25\% Resolution}
	\label{Table:computationtime}
		\begin{tabular}{l c c}
			\hline
			Methods  &25\%\ &100\%\\
			\hline
			ANTS			&08:32:24  &07:55:12 (affine only)\\
			
			Elastix 		&00:40:17   &27:31:00 (affine+deform)\\
			NiftyReg		&02:10:45   &Unable\\
			IRTK			&11:29:16  &Unable\\
                VoxelMorph      &00:00:55          &$00:07:00\approx$(deform)\\
                InvGAN          &00:01:10 &00:10:00$\approx$(deform)\\
			\hline
		\end{tabular}
	\end{center}
\end{table}

\section{Discussion}
In this paper we proposed a patch-based deep learning method for registration of very high resolution tissue cleared images. The proposed method is unsupervised and produced high registration accuracy on the tissue cleared dataset. Experimentation on two different resolution and on the anatomical landmarks indicates its ability as a general purpose registration method for large images in a resource constrained environment.
At 25\% resolution, the image dimension is $640\times540\times169$, which is still very high compared to other imaging modalities (MR,CT etc) and the quantitative  accuracy of the proposed method is the highest in terms of CC metric. In MI metric, the registration accuracy is very similar to ANTS. We also conducted the test on a non-GAN version of our proposed architecture. The non-GAN version achieves even higher accuracy among all the methods. We checked the qualitative results of non-GAN version and found that non-GAN version has serious box like artifacts due to patch based training which are removed by adversarial training. The high registration accuracy and good qualitative performance indicates the applicability and reliability of InvGAN on high resolution images.
\\ To further test its applicability we conduct training and evaluation at 100\% resolution. Training and testing at such extent required very expensive computational resources. Despite high resource consumption, the InvGAN achieves comparable accuracy where most of the baseline methods failed except the Elastix tool. The reason most of the conventional registration tools failed at 100\% resolution is the computational resources. We tried to apply those methods with 4TB of RAM (in a Big-data machine) and still they failed. On the other hand, proposed InvGAN achieves registration at 100\% with 64GB RAM and took only 9 to 10 mins. This experiment indicates that InvGAN is scale-able to any resolution with much less computational resources. 
\\ Anatomical landmark registration is considered a gold standard to validate performance of any registration method. Though, our InvGAN is an intensity based registration method, we decided to conduct landmark test for our method and other baseline methods. The performance of our method at 25\% and 100\% indicates its strengths as a holistic registration method while landmark test validates its performance on very local level. Small distance of landmarks after registration is a strong indicator of its performance at local level. 
\\A deficiency of our method as is common to most deep-learning based deformable registration methods is the requirement of affine alignment prior to the training and is work for future extension of this research.       
 
\section{Conclusion}
We developed a patch-based multi-decoder network for very high resolution tissue cleared image registration. Our proposed adversarial architecture has a generator and two discriminators. The discriminators examine the quality of the flow vectors and give their combined feedback to the generator network. Proper balancing is maintained in the loss function of the generator to make it learn. The quantitative and qualitative results of these analysis demonstrate that our approach outperformed multiple registration methods. Despite technical constraints and resource limitation, our method is trained and tested on 100\% resolution. The qualitative and quantitative performance at 100\% resolution of our method can be a reference point for future research in developing very high resolution biological image analysis pipeline.

\section*{CRediT authorship contribution statement}
\textbf{Abdullah Nazib:} Conceptualization, Methodology, Software, Formal analysis, Visualization, Writing – original draft, Writing – review \& editing.
\textbf{Clinton Fookes:} Conceptualization, Formal analysis, Writing – review \& editing. 
\textbf{Dimitri Perrin:} Conceptualization, Methodology, Formal analysis, Writing – review \& editing, Project administration.

\section*{Declaration of Competing Interest}
The authors declare that they have no known competing financial interests or personal relationships that could have appeared to influence the work reported in this paper.

\section*{Acknowledgments}
The work was supported by a QUT Postgraduate Research Award (A.N.).
Computational resources and services used in this work were provided by the eResearch Office, Queensland University of Technology, Brisbane, Australia.

\bibliographystyle{model1-num-names.bst}
\bibliography{references}

\begin{thebibliography}{28}
\expandafter\ifx\csname natexlab\endcsname\relax\def\natexlab#1{#1}\fi
\providecommand{\url}[1]{\texttt{#1}}
\providecommand{\href}[2]{#2}
\providecommand{\path}[1]{#1}
\providecommand{\DOIprefix}{doi:}
\providecommand{\ArXivprefix}{arXiv:}
\providecommand{\URLprefix}{URL: }
\providecommand{\Pubmedprefix}{pmid:}
\providecommand{\doi}[1]{\href{http://dx.doi.org/#1}{\path{#1}}}
\providecommand{\Pubmed}[1]{\href{pmid:#1}{\path{#1}}}
\providecommand{\bibinfo}[2]{#2}
\ifx\xfnm\relax \def\xfnm[#1]{\unskip,\space#1}\fi
\bibitem[{Dodt et~al.(2007)Dodt, Leischner, Schierloh, J{\"{a}}hrling, Mauch, Deininger, Deussing, Eder, Zieglg{\"{a}}nsberger, and Becker}]{Dodt2007}
\bibinfo{author}{H.-U. Dodt}, \bibinfo{author}{U.~Leischner}, \bibinfo{author}{A.~Schierloh}, \bibinfo{author}{N.~J{\"{a}}hrling}, \bibinfo{author}{C.~P. Mauch}, \bibinfo{author}{K.~Deininger}, \bibinfo{author}{J.~M. Deussing}, \bibinfo{author}{M.~Eder}, \bibinfo{author}{W.~Zieglg{\"{a}}nsberger}, \bibinfo{author}{K.~Becker},
\newblock \bibinfo{title}{{Ultramicroscopy: three-dimensional visualization of neuronal networks in the whole mouse brain}},
\newblock \bibinfo{journal}{Nature Methods} \bibinfo{volume}{4} (\bibinfo{year}{2007}) \bibinfo{pages}{331--336}.
\bibitem[{Hama et~al.(2011)Hama, Kurokawa, Kawano, Ando, Shimogori, Noda, Fukami, Sakaue-Sawano, and Miyawaki}]{Hama2011}
\bibinfo{author}{H.~Hama}, \bibinfo{author}{H.~Kurokawa}, \bibinfo{author}{H.~Kawano}, \bibinfo{author}{R.~Ando}, \bibinfo{author}{T.~Shimogori}, \bibinfo{author}{H.~Noda}, \bibinfo{author}{K.~Fukami}, \bibinfo{author}{A.~Sakaue-Sawano}, \bibinfo{author}{A.~Miyawaki},
\newblock \bibinfo{title}{{Scale: a chemical approach for fluorescence imaging and reconstruction of transparent mouse brain}},
\newblock \bibinfo{journal}{Nature Neuroscience} \bibinfo{volume}{14} (\bibinfo{year}{2011}) \bibinfo{pages}{1481--1488}.
\bibitem[{Ke et~al.(2013)Ke, Fujimoto, and Imai}]{Ke2013}
\bibinfo{author}{M.-T. Ke}, \bibinfo{author}{S.~Fujimoto}, \bibinfo{author}{T.~Imai},
\newblock \bibinfo{title}{{SeeDB: a simple and morphology-preserving optical clearing agent for neuronal circuit reconstruction}},
\newblock \bibinfo{journal}{Nature Neuroscience} \bibinfo{volume}{16} (\bibinfo{year}{2013}) \bibinfo{pages}{1154--1161}.
\bibitem[{Chung and Deisseroth(2013)}]{Chung2013}
\bibinfo{author}{K.~Chung}, \bibinfo{author}{K.~Deisseroth},
\newblock \bibinfo{title}{{CLARITY for mapping the nervous system}},
\newblock \bibinfo{journal}{Nature Methods} \bibinfo{volume}{10} (\bibinfo{year}{2013}) \bibinfo{pages}{508--513}.
\bibitem[{Renier et~al.(2014)Renier, Wu, Simon, Yang, Ariel, and Tessier-Lavigne}]{Renier2014}
\bibinfo{author}{N.~Renier}, \bibinfo{author}{Z.~Wu}, \bibinfo{author}{D.~J. Simon}, \bibinfo{author}{J.~Yang}, \bibinfo{author}{P.~Ariel}, \bibinfo{author}{M.~Tessier-Lavigne},
\newblock \bibinfo{title}{{IDISCO: A simple, rapid method to immunolabel large tissue samples for volume imaging}},
\newblock \bibinfo{journal}{Cell} \bibinfo{volume}{159} (\bibinfo{year}{2014}) \bibinfo{pages}{896--910}.
\bibitem[{Susaki et~al.(2014)Susaki, Tainaka, Perrin, Kishino, Tawara, Watanabe, Yokoyama, Onoe, Eguchi, Yamaguchi, Abe, Kiyonari, Shimizu, Miyawaki, Yokota, and Ueda}]{Susaki2014}
\bibinfo{author}{E.~A. Susaki}, \bibinfo{author}{K.~Tainaka}, \bibinfo{author}{D.~Perrin}, \bibinfo{author}{F.~Kishino}, \bibinfo{author}{T.~Tawara}, \bibinfo{author}{T.~M. Watanabe}, \bibinfo{author}{C.~Yokoyama}, \bibinfo{author}{H.~Onoe}, \bibinfo{author}{M.~Eguchi}, \bibinfo{author}{S.~Yamaguchi}, \bibinfo{author}{T.~Abe}, \bibinfo{author}{H.~Kiyonari}, \bibinfo{author}{Y.~Shimizu}, \bibinfo{author}{A.~Miyawaki}, \bibinfo{author}{H.~Yokota}, \bibinfo{author}{H.~R. Ueda},
\newblock \bibinfo{title}{{Whole-brain imaging with single-cell resolution using chemical cocktails and computational analysis}},
\newblock \bibinfo{journal}{Cell} \bibinfo{volume}{157} (\bibinfo{year}{2014}) \bibinfo{pages}{726--739}.
\bibitem[{Balakrishnan et~al.(2018)Balakrishnan, Zhao, Sabuncu, Dalca, and Guttag}]{Balakrishnan2018}
\bibinfo{author}{G.~Balakrishnan}, \bibinfo{author}{A.~Zhao}, \bibinfo{author}{M.~R. Sabuncu}, \bibinfo{author}{A.~V. Dalca}, \bibinfo{author}{J.~Guttag},
\newblock \bibinfo{title}{{An Unsupervised Learning Model for Deformable Medical Image Registration}},
\newblock in: \bibinfo{booktitle}{2018 IEEE/CVF Conference on Computer Vision and Pattern Recognition}, \bibinfo{publisher}{IEEE}, \bibinfo{year}{2018}, pp. \bibinfo{pages}{9252--9260}. \URLprefix \url{https://ieeexplore.ieee.org/document/8579062/}. \DOIprefix\doi{10.1109/CVPR.2018.00964}.
\bibitem[{Eppenhof and Pluim(2019)}]{Eppenhof2019}
\bibinfo{author}{K.~A.~J. Eppenhof}, \bibinfo{author}{J.~P.~W. Pluim},
\newblock \bibinfo{title}{{Pulmonary CT Registration Through Supervised Learning With Convolutional Neural Networks}},
\newblock \bibinfo{journal}{IEEE Transactions on Medical Imaging} \bibinfo{volume}{38} (\bibinfo{year}{2019}) \bibinfo{pages}{1097--1105}.
\bibitem[{Cao et~al.(2018)Cao, Yang, Zhang, Wang, Yap, and Shen}]{Cao2018}
\bibinfo{author}{X.~Cao}, \bibinfo{author}{J.~Yang}, \bibinfo{author}{J.~Zhang}, \bibinfo{author}{Q.~Wang}, \bibinfo{author}{P.-T. Yap}, \bibinfo{author}{D.~Shen},
\newblock \bibinfo{title}{{Deformable Image Registration Using a Cue-Aware Deep Regression Network}},
\newblock \bibinfo{journal}{IEEE Transactions on Biomedical Engineering} \bibinfo{volume}{65} (\bibinfo{year}{2018}) \bibinfo{pages}{1900--1911}.
\bibitem[{Yang et~al.(2016)Yang, Kwitt, and Niethammer}]{Yang2016}
\bibinfo{author}{X.~Yang}, \bibinfo{author}{R.~Kwitt}, \bibinfo{author}{M.~Niethammer},
\newblock \bibinfo{title}{Fast predictive image registration},
\newblock in: \bibinfo{editor}{G.~Carneiro}, \bibinfo{editor}{D.~Mateus}, \bibinfo{editor}{L.~Peter}, \bibinfo{editor}{A.~Bradley}, \bibinfo{editor}{J.~M. R.~S. Tavares}, \bibinfo{editor}{V.~Belagiannis}, \bibinfo{editor}{J.~P. Papa}, \bibinfo{editor}{J.~C. Nascimento}, \bibinfo{editor}{M.~Loog}, \bibinfo{editor}{Z.~Lu}, \bibinfo{editor}{J.~S. Cardoso}, \bibinfo{editor}{J.~Cornebise} (Eds.), \bibinfo{booktitle}{Deep Learning and Data Labeling for Medical Applications}, \bibinfo{publisher}{Springer International Publishing}, \bibinfo{address}{Cham}, \bibinfo{year}{2016}, pp. \bibinfo{pages}{48--57}.
\bibitem[{Roh{\'{e}} et~al.(2017)Roh{\'{e}}, Datar, Heimann, Sermesant, and Pennec}]{Rohe2017}
\bibinfo{author}{M.-M. Roh{\'{e}}}, \bibinfo{author}{M.~Datar}, \bibinfo{author}{T.~Heimann}, \bibinfo{author}{M.~Sermesant}, \bibinfo{author}{X.~Pennec},
\newblock \bibinfo{title}{{SVF-Net: Learning Deformable Image Registration Using Shape Matching}},
\newblock in: \bibinfo{editor}{M.~Descoteaux}, \bibinfo{editor}{L.~Maier-Hein}, \bibinfo{editor}{A.~Franz}, \bibinfo{editor}{P.~Jannin}, \bibinfo{editor}{D.~L. Collins}, \bibinfo{editor}{S.~Duchesne} (Eds.), \bibinfo{booktitle}{Medical Image Computing and Computer Assisted Intervention − MICCAI 2017}, \bibinfo{publisher}{Springer International Publishing}, \bibinfo{address}{Cham}, \bibinfo{year}{2017}, pp. \bibinfo{pages}{266--274}.
\bibitem[{Sokooti et~al.(2017)Sokooti, de~Vos, Berendsen, Lelieveldt, I{\v{s}}gum, and Staring}]{Sokooti2017}
\bibinfo{author}{H.~Sokooti}, \bibinfo{author}{B.~de~Vos}, \bibinfo{author}{F.~Berendsen}, \bibinfo{author}{B.~P.~F. Lelieveldt}, \bibinfo{author}{I.~I{\v{s}}gum}, \bibinfo{author}{M.~Staring},
\newblock \bibinfo{title}{{Nonrigid Image Registration Using Multi-scale 3D Convolutional Neural Networks}},
\newblock in: \bibinfo{editor}{M.~Descoteaux}, \bibinfo{editor}{L.~Maier-Hein}, \bibinfo{editor}{A.~Franz}, \bibinfo{editor}{P.~Jannin}, \bibinfo{editor}{D.~L. Collins}, \bibinfo{editor}{S.~Duchesne} (Eds.), \bibinfo{booktitle}{Medical Image Computing and Computer Assisted Intervention − MICCAI 2017}, \bibinfo{publisher}{Springer International Publishing}, \bibinfo{address}{Cham}, \bibinfo{year}{2017}, pp. \bibinfo{pages}{232--239}.
\bibitem[{Li and Fan(2018)}]{Li2018}
\bibinfo{author}{H.~Li}, \bibinfo{author}{Y.~Fan},
\newblock \bibinfo{title}{{Non-Rigid Image Registration Using Self-Supervised Fully Convolutional Networks without Training Data}},
\newblock \bibinfo{journal}{arXiv preprint arXiv:1801.04012}  (\bibinfo{year}{2018}).
\bibitem[{Yan et~al.(2018)Yan, Xu, Rastinehad, and Wood}]{Yan2018}
\bibinfo{author}{P.~Yan}, \bibinfo{author}{S.~Xu}, \bibinfo{author}{A.~R. Rastinehad}, \bibinfo{author}{B.~J. Wood},
\newblock \bibinfo{title}{Adversarial image registration with application for mr and trus image fusion},
\newblock in: \bibinfo{booktitle}{MLMI@MICCAI}, \bibinfo{year}{2018}.
\bibitem[{Arjovsky et~al.(2017)Arjovsky, Chintala, and Bottou}]{Arjovsky2017}
\bibinfo{author}{M.~Arjovsky}, \bibinfo{author}{S.~Chintala}, \bibinfo{author}{L.~Bottou},
\newblock \bibinfo{title}{{W}asserstein generative adversarial networks},
\newblock in: \bibinfo{editor}{D.~Precup}, \bibinfo{editor}{Y.~W. Teh} (Eds.), \bibinfo{booktitle}{Proceedings of the 34th International Conference on Machine Learning}, volume~\bibinfo{volume}{70} of \textit{\bibinfo{series}{Proceedings of Machine Learning Research}}, \bibinfo{publisher}{PMLR}, \bibinfo{year}{2017}, pp. \bibinfo{pages}{214--223}. \URLprefix \url{https://proceedings.mlr.press/v70/arjovsky17a.html}.
\bibitem[{Mahapatra et~al.(2018)Mahapatra, Antony, Sedai, and Garnavi}]{Mahapatra2018a}
\bibinfo{author}{D.~Mahapatra}, \bibinfo{author}{B.~Antony}, \bibinfo{author}{S.~Sedai}, \bibinfo{author}{R.~Garnavi},
\newblock \bibinfo{title}{Deformable medical image registration using generative adversarial networks},
\newblock in: \bibinfo{booktitle}{2018 IEEE 15th International Symposium on Biomedical Imaging (ISBI 2018)}, \bibinfo{year}{2018}, pp. \bibinfo{pages}{1449--1453}. \DOIprefix\doi{10.1109/ISBI.2018.8363845}.
\bibitem[{Zhu et~al.(2017)Zhu, Park, Isola, and Efros}]{Zhu2018}
\bibinfo{author}{J.-Y. Zhu}, \bibinfo{author}{T.~Park}, \bibinfo{author}{P.~Isola}, \bibinfo{author}{A.~A. Efros},
\newblock \bibinfo{title}{Unpaired image-to-image translation using cycle-consistent adversarial networks},
\newblock in: \bibinfo{booktitle}{Computer Vision (ICCV), 2017 IEEE International Conference on}, \bibinfo{year}{2017}.
\bibitem[{de~Vos et~al.(2017)de~Vos, Berendsen, Viergever, Staring, and I{\v{s}}gum}]{DeVos2017}
\bibinfo{author}{B.~D. de~Vos}, \bibinfo{author}{F.~F. Berendsen}, \bibinfo{author}{M.~A. Viergever}, \bibinfo{author}{M.~Staring}, \bibinfo{author}{I.~I{\v{s}}gum},
\newblock \bibinfo{title}{{End-to-end unsupervised deformable image registration with a convolutional neural network}},
\newblock in: \bibinfo{booktitle}{Lecture Notes in Computer Science (including subseries Lecture Notes in Artificial Intelligence and Lecture Notes in Bioinformatics)}, volume \bibinfo{volume}{10553 LNCS}, \bibinfo{publisher}{Springer Verlag}, \bibinfo{year}{2017}, pp. \bibinfo{pages}{204--212}. \DOIprefix\doi{10.1007/978-3-319-67558-9_24}. \href{http://arxiv.org/abs/1704.06065}{\tt arXiv:1704.06065}.
\bibitem[{Klein et~al.(2010)Klein, Staring, Murphy, Viergever, and Pluim}]{Klein2010a}
\bibinfo{author}{S.~Klein}, \bibinfo{author}{M.~Staring}, \bibinfo{author}{K.~Murphy}, \bibinfo{author}{M.~A. Viergever}, \bibinfo{author}{J.~P. Pluim},
\newblock \bibinfo{title}{{Elastix: A toolbox for intensity-based medical image registration}},
\newblock \bibinfo{journal}{IEEE Transactions on Medical Imaging} \bibinfo{volume}{29} (\bibinfo{year}{2010}) \bibinfo{pages}{196--205}.
\bibitem[{Fan et~al.(2018)Fan, Cao, Xue, Yap, and Shen}]{Fan2018}
\bibinfo{author}{J.~Fan}, \bibinfo{author}{X.~Cao}, \bibinfo{author}{Z.~Xue}, \bibinfo{author}{P.-T. Yap}, \bibinfo{author}{D.~Shen},
\newblock \bibinfo{title}{{Adversarial Similarity Network for Evaluating Image Alignment in Deep Learning based Registration.}},
\newblock \bibinfo{journal}{Medical image computing and computer-assisted intervention : MICCAI ... International Conference on Medical Image Computing and Computer-Assisted Intervention} \bibinfo{volume}{11070} (\bibinfo{year}{2018}) \bibinfo{pages}{739--746}.
\bibitem[{Ronneberger et~al.(2015)Ronneberger, Fischer, and Brox}]{Ronneberger2015}
\bibinfo{author}{O.~Ronneberger}, \bibinfo{author}{P.~Fischer}, \bibinfo{author}{T.~Brox},
\newblock \bibinfo{title}{{U-Net: Convolutional Networks for Biomedical Image Segmentation}},
\newblock \bibinfo{journal}{arXiv preprint} \bibinfo{volume}{abs/1505.0} (\bibinfo{year}{2015}) \bibinfo{pages}{234--241}.
\bibitem[{Susaki et~al.(2015)Susaki, Tainaka, Perrin, Yukinaga, Kuno, and Ueda}]{Susaki2015}
\bibinfo{author}{E.~A. Susaki}, \bibinfo{author}{K.~Tainaka}, \bibinfo{author}{D.~Perrin}, \bibinfo{author}{H.~Yukinaga}, \bibinfo{author}{A.~Kuno}, \bibinfo{author}{H.~R. Ueda},
\newblock \bibinfo{title}{{Advanced CUBIC protocols for whole-brain and whole-body clearing and imaging}},
\newblock \bibinfo{journal}{Nature Protocols} \bibinfo{volume}{10} (\bibinfo{year}{2015}) \bibinfo{pages}{1709--1727}.
\bibitem[{Xu et~al.(2016)Xu, Lee, Heinrich, Modat, Rueckert, Ourselin, Abramson, and Landman}]{Xu2016}
\bibinfo{author}{Z.~Xu}, \bibinfo{author}{C.~P. Lee}, \bibinfo{author}{M.~P. Heinrich}, \bibinfo{author}{M.~Modat}, \bibinfo{author}{D.~Rueckert}, \bibinfo{author}{S.~Ourselin}, \bibinfo{author}{R.~G. Abramson}, \bibinfo{author}{B.~A. Landman},
\newblock \bibinfo{title}{{Evaluation of Six Registration Methods for the Human Abdomen on Clinically Acquired CT}},
\newblock \bibinfo{journal}{IEEE Transactions on Biomedical Engineering} \bibinfo{volume}{63} (\bibinfo{year}{2016}) \bibinfo{pages}{1563--1572}.
\bibitem[{Rueckert et~al.(1999)Rueckert, Sonoda, Hayes, Hill, Leach, and Hawkes}]{Rueckert1999}
\bibinfo{author}{D.~Rueckert}, \bibinfo{author}{L.~Sonoda}, \bibinfo{author}{C.~Hayes}, \bibinfo{author}{D.~Hill}, \bibinfo{author}{M.~Leach}, \bibinfo{author}{D.~Hawkes},
\newblock \bibinfo{title}{{Nonrigid registration using free-form deformations: application to breast MR images}},
\newblock \bibinfo{journal}{IEEE Transactions on Medical Imaging} \bibinfo{volume}{18} (\bibinfo{year}{1999}) \bibinfo{pages}{712--721}.
\bibitem[{Hammelrath et~al.(2016)Hammelrath, {\v{S}}koki{\'{c}}, Khmelinskii, Hess, van~der Knaap, Staring, Lelieveldt, Wiedermann, and Hoehn}]{Hammelrath2016}
\bibinfo{author}{L.~Hammelrath}, \bibinfo{author}{S.~{\v{S}}koki{\'{c}}}, \bibinfo{author}{A.~Khmelinskii}, \bibinfo{author}{A.~Hess}, \bibinfo{author}{N.~van~der Knaap}, \bibinfo{author}{M.~Staring}, \bibinfo{author}{B.~P. Lelieveldt}, \bibinfo{author}{D.~Wiedermann}, \bibinfo{author}{M.~Hoehn},
\newblock \bibinfo{title}{{Morphological maturation of the mouse brain: An in vivo MRI and histology investigation}},
\newblock \bibinfo{journal}{NeuroImage} \bibinfo{volume}{125} (\bibinfo{year}{2016}) \bibinfo{pages}{144--152}.
\bibitem[{Avants et~al.(2008)Avants, Epstein, Grossman, and Gee}]{Avants2008}
\bibinfo{author}{B.~Avants}, \bibinfo{author}{C.~Epstein}, \bibinfo{author}{M.~Grossman}, \bibinfo{author}{J.~Gee},
\newblock \bibinfo{title}{{Symmetric diffeomorphic image registration with cross-correlation: Evaluating automated labeling of elderly and neurodegenerative brain}},
\newblock \bibinfo{journal}{Medical Image Analysis} \bibinfo{volume}{12} (\bibinfo{year}{2008}) \bibinfo{pages}{26--41}.
\bibitem[{Avants et~al.(2011)Avants, Tustison, Song, Cook, Klein, and Gee}]{Avants2011}
\bibinfo{author}{B.~B. Avants}, \bibinfo{author}{N.~J. Tustison}, \bibinfo{author}{G.~Song}, \bibinfo{author}{P.~A. Cook}, \bibinfo{author}{A.~Klein}, \bibinfo{author}{J.~C. Gee},
\newblock \bibinfo{title}{{A reproducible evaluation of ANTs similarity metric performance in brain image registration}},
\newblock \bibinfo{journal}{NeuroImage} \bibinfo{volume}{54} (\bibinfo{year}{2011}) \bibinfo{pages}{2033--2044}.
\bibitem[{Modat et~al.(2010)Modat, Ridgway, Taylor, Lehmann, Barnes, Hawkes, Fox, and Ourselin}]{Modat2010}
\bibinfo{author}{M.~Modat}, \bibinfo{author}{G.~R. Ridgway}, \bibinfo{author}{Z.~A. Taylor}, \bibinfo{author}{M.~Lehmann}, \bibinfo{author}{J.~Barnes}, \bibinfo{author}{D.~J. Hawkes}, \bibinfo{author}{N.~C. Fox}, \bibinfo{author}{S.~Ourselin},
\newblock \bibinfo{title}{{Fast free-form deformation using graphics processing units}},
\newblock \bibinfo{journal}{Computer Methods and Programs in Biomedicine} \bibinfo{volume}{98} (\bibinfo{year}{2010}) \bibinfo{pages}{278--284}.

\end{thebibliography}
\end{document}